\documentclass{emulateapj-rtx4}
\usepackage{lscape}

\submitted{Accepted for Publication in the Astronomical Journal: July 4, 2014}
\shorttitle{Bulge RGB Abundance Trends}
\shortauthors{Johnson et al.}

\newcommand\iso[2]{$^{\rm #1}$#2}

\begin{document}

\title{Light, Alpha, and Fe--Peak Element Abundances in the Galactic Bulge}

\author{
Christian I. Johnson\altaffilmark{1,2},
R. Michael Rich\altaffilmark{3},
Chiaki Kobayashi\altaffilmark{4},
Andrea Kunder\altaffilmark{5}, and
Andreas Koch\altaffilmark{6}
}

\altaffiltext{1}{Harvard--Smithsonian Center for Astrophysics, 60 Garden
Street, MS--15, Cambridge, MA 02138, USA; cjohnson@cfa.harvard.edu}

\altaffiltext{2}{Clay Fellow}

\altaffiltext{3}{Department of Physics and Astronomy, UCLA, 430 Portola Plaza,
Box 951547, Los Angeles, CA 90095-1547, USA; rmr@astro.ucla.edu}

\altaffiltext{4}{Centre for Astrophysics Research, University of Hertfordshire,
Hatfield  AL10 9AB, UK; c.kobayashi@herts.ac.uk}

\altaffiltext{5}{Leibniz--Institute f\"{u}r Astrophysik Potsdam (AIP), Ander
Sternwarte 16, D--14482, Potsdam, Germany; akunder@aip.de}

\altaffiltext{6}{Zentrum f\"{u}r Astronomie der Universit\"{a}t
Heidelberg, Landessternwarte, K\"{o}nigstuhl 12, Heidelberg, Germany;
akoch@lsw.uni-heidelberg.de}

\begin{abstract}

We present radial velocities and chemical abundances of O, Na, Mg, Al, Si, Ca,
Cr, Fe, Co, Ni, and Cu for a sample of 156 red giant branch stars in two
Galactic bulge fields centered near (l,b)=($+$5.25,--3.02) and (0,--12).  The
($+$5.25,--3.02) field also includes observations of the bulge globular cluster
NGC 6553.  The results are based on high resolution (R$\sim$20,000), high 
signal--to--noise (S/N$\ga$70) FLAMES--GIRAFFE spectra obtained through the 
ESO archive.  However, we only selected a subset of the original observations
that included spectra with both high S/N and that did not show strong
TiO absorption bands.  The present work extends previous analyses of this data
set beyond Fe and the $\alpha$--elements Mg, Si, Ca, and Ti.  While we find 
reasonable agreement with past work, the data presented here indicate that the
bulge may exhibit a different chemical composition than the local thick disk, 
especially at [Fe/H]$\ga$--0.5.  In particular, the bulge [$\alpha$/Fe] ratios
may remain enhanced to a slightly higher [Fe/H] than the thick disk and the 
Fe--peak elements Co, Ni, and Cu appear enhanced compared to the disk.  There
is also some evidence that the [Na/Fe] (but not [Al/Fe]) trends between the 
bulge and local disk may be different at low and high metallicity.  We also
find that the velocity dispersion decreases as a function of increasing [Fe/H] 
for both fields, and do not detect any significant cold, high velocity 
population.  A comparison with chemical enrichment models indicates that a 
significant fraction of hypernovae are required to explain the bulge abundance 
trends, and that initial mass functions that are steep, top--heavy (and do not 
include strong outflow), or truncated to avoid including contributions from 
stars $>$40 M$_{\rm \odot}$ are ruled out, in particular because of 
disagreement with the Fe--peak abundance data.  For most elements, the NGC 6553
stars exhibit nearly identical abundance trends to comparable metallicity bulge
field stars.  However, the star--to--star scatter and mean [Na/Fe] ratios 
appear higher in the cluster, perhaps indicating additional self--enrichment.

\end{abstract}

\keywords{stars: abundances, Galactic bulge: general, bulge:
Galaxy: bulge, stars: Population II}

\section{INTRODUCTION}

Understanding the formation and subsequent evolution of the Galactic bulge is
important both for interpreting observations of extragalactic populations and 
for constraining Galaxy chemodynamical formation models.  Recent large sample 
spectroscopic surveys, such as the Bulge Radial Velocity Assay (BRAVA; Rich et 
al. 2007a; Howard et al. 2008; Howard et al. 2009; Kunder et al. 2012), the
Abundances and Radial velocity Galactic Origins Survey (ARGOS; Freeman et al.
2013; Ness et al. 2012; 2013b), the Apache Point Observatory Galactic 
Evolution Experiment (APOGEE; Majewski et al. 2010; Nidever et al. 2012), and 
the GIRAFFE Inner Bulge Survey (GIBS; Zoccali et al. 2014) provide a coherent 
view of the bulge as a barred, triaxial system exhibiting cylindrical 
rotation.  Photometric and star count studies have also discovered 
a double red clump along some bulge sight lines that traces out an X--shaped
structure (McWilliam \& Zoccali 2010; Nataf et al. 2010; Saito et al. 2011).
This structure appears to be dominated by stars with [Fe/H]$>$--0.5 on 
bar--supporting orbits (Soto et al. 2007; Babusiaux et al. 2010; Ness et al. 
2012; Uttenthaler et al. 2012; but see also Nataf et al. 2014).

Inclusive with these data are detailed composition analyses of field stars from
moderate and high resolution spectroscopy (McWilliam \& Rich 1994; 
Ram{\'{\i}}rez et al. 2000; Rich \& Origlia 2005; Cunha \& Smith 2006; 
Fulbright et al. 2006; Zoccali et al. 2006; Fulbright et al. 2007; Lecureur et 
al. 2007; Rich et al. 2007b; Cunha et al. 2008; Mel{\'e}ndez et al. 2008; 
Zoccali et al. 2008; Alves--Brito et al. 2010; Bensby et al. 2010a; Ryde et al.
2010; Bensby et al. 2011; Gonzalez et al. 2011; Hill et al. 2011; Johnson et 
al. 2011; Johnson et al. 2012; Rich et al. 2012; Uttenthaler et al. 2012; 
Barbuy et al. 2013; Bensby et al. 2013; Garc{\'{\i}}a P{\'e}rez et al. 2013; 
Johnson et al. 2013a; Ness et al. 2013a; J{\"o}nsson et al. 2014) finding, at 
least in a general sense, that the bulge is composed of stars spanning more 
than a factor of 100 in [Fe/H]\footnote{[A/B]$\equiv$log(N$_{\rm A}$/N$_{\rm B}$)$_{\rm star}$--log(N$_{\rm A}$/N$_{\rm B}$)$_{\sun}$ and log $\epsilon$(A)$\equiv$log(N$_{\rm A}$/N$_{\rm H}$)+12.0 for elements A and B.}, that bulge stars
are uniformly enhanced in their [$\alpha$/Fe] ratios at low metallicity 
relative to the thin disk, and that the median [Fe/H] along bulge sight lines
decreases as a function of increasing Galactic latitude (i.e., there is a 
metallicity gradient).  The enhanced [$\alpha$/Fe] abundances, coupled with the
low [La/Eu] ratios of bulge stars (McWilliam et al. 2010; Johnson et al. 2012),
are consistent with the notion that the bulge formed rapidly ($\la$1--3 Gyr).  
In fact, the bulge appears uniformly old ($\sim$10 Gyr) in age studies based on
color--magnitude diagram analyses (e.g., Ortolani et al. 1995; Zoccali et al. 
2003; Clarkson et al. 2008; Valenti et al. 2013; but see also Ness et al.
2014), and Clarkson et al. (2011) estimate from the blue straggler population 
in an inner bulge field that a truly young ($<$5 Gyr) population should not 
constitute more than $\sim$3.4$\%$ of the bulge.  In contrast, ages derived 
from microlensed dwarf studies (e.g., Bensby et al. 2013) find that while all 
metal--poor bulge stars are uniformly old, $\sim$5--25$\%$ of metal--rich 
stars, at least near the Galactic plane, may be only $\sim$2--8 Gyr in age.  

While the observational data continue to grow, the difficult task of assembling
the pieces into a fully self--consistent model of the bulge's formation 
remains open.  The chemodynamical bulge data are challenging to interpret. 
The bulge's predominantly old age, enhanced [$\alpha$/Fe] ratios, vertical
metallicity gradient, and the existence of possible ``primordial building
blocks" such as Terzan 5 (e.g., Ferraro et al 2009; Origlia et al. 2011; 2013)
are more consistent with the classical, merger built formation scenario.
However, the bulge's boxy X--shape, similar composition characteristics to at 
least the thick disk, and cylindrical rotation profile suggest that the bulge
formed via secular evolution from a buckling disk instability and may
be a ``pseudobulge" (e.g., Kormendy \& Kennicutt 2004; but see also 
Zoccali et al. 2014).  While Shen et al. (2010) rule out a classical bulge
component that exceeds $\sim$8$\%$ of the disk mass, it may still be possible
for a bar to form within a pre--existing classical bulge (e.g., Saha et al.
2012).  Additionally, evidence such as the metallicity gradient may not 
be unique to the classical bulge scenario, and may be consistent with a 
secular evolution model in which a radial metallicity gradient in the buckling
disk is transformed into a vertical gradient in the resultant bar 
(Martinez--Valpuesta \& Gerhard 2013).  The bulge may also be composed of at 
least two stellar populations with different composition and kinematics 
(Babusiaux et al. 2010; Hill et al. 2011; Bensby et al. 2011; Bensby et al. 
2013; Ness et al. 2013a).  However, at the moment the exact nature of these 
potentially distinct stellar populations is far from certain.

Although most of the chemical abundance work mentioned previously has 
focused on the [Fe/H] and [$\alpha$/Fe] ratios in comparison with the thin
and thick disks, the light odd--Z and Fe--peak (and also neutron--capture) 
elements also provide discriminatory power between models and other stellar 
populations (e.g., see Kobayashi et al. 2011; their Figure 14).  The Fe--peak 
elements in particular are useful as they may be sensitive to 
formation environment and metallicity.  For example, the metallicity
dependent yields and increased contributions of massive stars are predicted to 
produce enhanced [Cu/Fe] and [Zn/Fe] ratios in the bulge compared to the local 
disk.  Similarly, if the bulge formed with a significantly flatter initial mass
function (IMF) than the disk then bulge stars should exhibit very large [Co/Fe]
and [Zn/Fe] ratios (Nomoto et al. 2013).  Therefore, here we measure abundances
of the Fe--peak elements Cr, Fe, Co, Ni, and Cu, in addition to the light 
odd--Z and $\alpha$--elements O, Na, Mg, Al, Si, and Ca, in 156 red giant 
branch (RGB) stars in two Galactic bulge fields at (l,b)=($+$5.25,--3.02) and 
(0,--12), and compare the abundance ratios with other bulge fields, the 
Galactic disk, and chemical enrichment models.

\section{OBSERVATIONS, TARGET SELECTION, AND DATA REDUCTION}

The FLAMES--GIRAFFE spectra for this project are based on data obtained from 
the ESO Science Archive Facility under request number 51251, which are based 
on observations collected at the European Southern Observatory, Paranal, Chile 
(ESO Program 073.B--0074).  Details regarding the selection of targets and 
input parameters (e.g., photometry and astrometry) are given in Zoccali et al.
(2008).  To briefly summarize, fibers were placed on K giants approximately
1--2 magnitudes brighter than the bulge red clump, and the spectra were obtained
in high resolution mode (R$\equiv$$\lambda$/$\Delta$$\lambda$$\sim$20,000).  
The original program by Zoccali et al. (2008) included four fields centered at 
(l,b)=($+$1.14,--4.18), ($+$0.21,--6.02), (0,--12), and ($+$5.25,--3.02).  
While the ($+$1.14,--4.18) and ($+$0.21,--6.02) fields were observed in the HR 
13, HR 14, and HR 15 setups (spanning $\sim$6100--6950 \AA), the
($+$5.25,--3.02) and (0,--12) fields were observed in the HR 11, HR 13, and 
HR 15 setups (5590--5835 \AA; 6100--6400 \AA; 6600--6950 \AA).  Since the HR 11
setup is the only one containing measurable copper lines, we have only analyzed 
GIRAFFE spectra from the ($+$5.25,--3.02) and (0,--12) fields.  We note that
the ($+$5.25,--3.02) field also includes the bulge globular cluster NGC 6553.

\begin{figure}
\epsscale{1.00}
\plotone{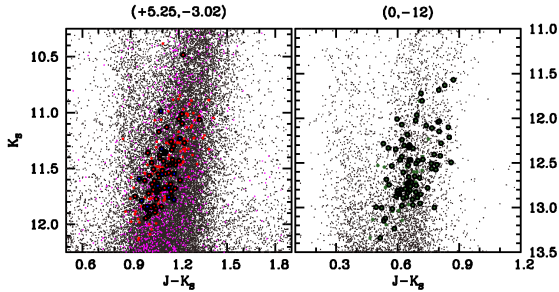}
\caption{\emph{left panel:} Color--magnitude diagram for the field centered
near (l,b)=($+$5.25,--3.02).  The filled red circles are all stars observed
with the FLAMES instrument.  The black outlined circles are those presented
in this paper.  The filled blue boxes indicate stars with radial velocities
and metallicities consistent with belonging to the globular cluster NGC 6553.
The small filled gray circles indicate all stars in the 2MASS catalog within
30$\arcmin$ of the central coordinates.  Similarly, the small filled magenta
circles indicate all stars in the 2MASS catalog within 5$\arcmin$ of NGC 6553.
\emph{right panel:} A similar plot but with the observed stars for the
(l,b)=(0,--12) field shown in green.}
\label{f1}
\end{figure}

Figure \ref{f1} shows a 2MASS (Skrutskie et al. 2006) color--magnitude diagram 
of our final target selection from the archival data.  The raw data set 
obtained from the ESO archive included observations of 205 RGB stars in the 
($+$5.25,--3.02) field and 109 RGB stars in the (0,--12) field.  However, we 
only analyzed spectra for which the co--added signal--to--noise (S/N) ratio 
exceeded $\sim$70.  We also discarded spectra that exhibited strong TiO 
absorption bands, for which a ``standard" equivalent width (EW) analysis would 
be inappropriate.  The final sample utilized here includes 75/205 stars 
(37$\%$) in the ($+$5.25,--3.02) field and 81/109 stars 
(74$\%$) in the (0,--12) field.  In Figure \ref{f1} we also identify stars 
that are likely members of NGC 6553 (see Section 3.5).  In particular, note 
the broad dispersion in the color--magnitude diagram of cluster members, as 
well as with stars within 5$\arcmin$ of the cluster center.  This highlights 
the combined effects of differential reddening and population mixing along the
line--of--sight toward the ($+$5.25,--3.02) field.  The star names and 
coordinates from the raw image headers and Zoccali et al. (2008), as well as
available 2MASS photometry and star identifiers, are provided in Table 1.

The raw science and calibration data were downloaded and re--reduced using
the GIRAFFE Base--Line Data Reduction Software (girBLDRS)\footnote{The girBLDRS
software can be downloaded at: http://girbldrs.sourceforge.net/.}.  In 
particular, the pipeline software was used to carry--out bias subtraction and 
overscan trimming, dark correction, fiber identification, flat--fielding, 
wavelength calibration, scattered light correction, and spectrum extraction.  
Sky subtraction was carried out using the IRAF\footnote{IRAF is distributed by 
the National Optical Astronomy Observatory, which is operated by the 
Association of Universities for Research in Astronomy, Inc., under cooperative 
agreement with the National Science Foundation.} \emph{skysub} routine.  
Individual exposures were continuum normalized using a low order polynomial 
via the IRAF \emph{continuum} routine, and the telluric band in the HR 13 
spectra was removed using the IRAF task \emph{telluric} and a set of FLAMES 
templates obtained during a different observing program with the same 
spectrograph setup.  The individual spectra were shifted to a common velocity 
scale (i.e., the heliocentric velocity was removed) and co--added using 
IRAF's \emph{scombine} task.

\section{DATA ANALYSIS}

\subsection{Model Stellar Atmospheres}

The four primary model atmosphere input parameters of effective temperature
(T$_{\rm eff}$), surface gravity (log(g)), metallicity ([Fe/H]), and 
microturbulence (vt) were determined via spectroscopic analyses.  For stars in
the (0,--12) field, we used the model parameters given in Zoccali et al.
(2008) as a starting point before converging to a solution.  However, the
adopted model atmosphere parameters for stars in the ($+$5.25,--3.02) field
are not provided in Zoccali et al. (2008) nor Gonzalez et al. (2011).  
Therefore, we adopted the generic values T$_{\rm eff}$=4500 K, log(g)=$+$2.0
cgs, [Fe/H]=--0.20 dex, and vt=1.5 km s$^{\rm -1}$ before converging to a 
solution.  The final parameters given in Table 1 were derived by enforcing Fe I
excitation equilibrium for T$_{\rm eff}$, ionization equilibrium between 
Fe I/II\footnote{For stars in which Fe II lines were not measurable, we 
adopted the average T$_{\rm eff}$/log(g) combination for other stars of 
comparable T$_{\rm eff}$ and [Fe/H].} for log(g), and removing trends in Fe I 
abundance versus line strength for vt.  The final models were interpolated 
within the available grid of AODFNEW ($\alpha$--enhanced) and ODFNEW 
(scaled--solar) ATLAS9 model atmospheres\footnote{The model atmosphere grid 
can be accessed at: http://wwwuser.oat.ts.astro.it/castelli/grids.html.} 
(Castelli \& Kurucz 2004).  Stars with [$\alpha$/Fe]$>$$+$0.15 were measured
using the $\alpha$--enhanced models, and we used the scaled--solar models for
stars with [$\alpha$/Fe]$<$$+$0.15.  However, the issue of an 
$\alpha$--enhanced versus scaled--solar model should not introduce an error in 
the abundance ratios that exceeds the $\sim$0.05--0.10 dex level (e.g., 
Fulbright et al. 2006; Alves--Brito et al. 2010; Johnson et al. 2013a).

\begin{figure}
\epsscale{1.00}
\plotone{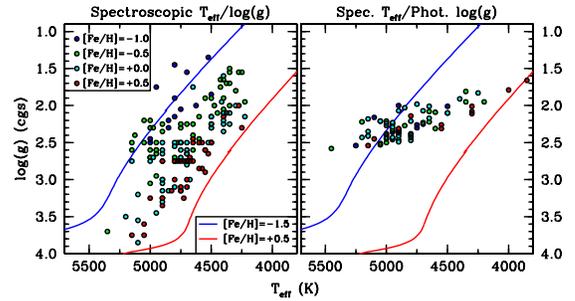}
\caption{\emph{left panel:} A plot of surface gravity versus effective
temperature for all stars analyzed in this paper.  The symbols are color--coded
into rough metallicity bins.  Metal--poor, $\alpha$--enhanced (blue) and
metal--rich, $\alpha$--normal (red) 10 Gyr isochrone sequences (Dotter et al.
2008) are shown for guidance.  \emph{right panel:} The effective temperature
(excitation equilibrium) and surface gravity (photometric) values employed by
Zoccali et al. (2008) for the same stars presented here in the (0,--12) field.
The literature model atmosphere parameters for stars in the ($+$5.25,--3.02)
field are not available for comparison.}
\label{f2}
\end{figure}

Figure \ref{f2} shows our spectroscopically determined temperature and surface 
gravity values in comparison with the the spectroscopic T$_{\rm eff}$ and 
photometric log(g) values given in Zoccali et al. (2008).  As is evident in
Figure \ref{f2}, the spectroscopic determination of both parameters leads to 
a more extended distribution of surface gravities.  This is likely due to the 
unavoidable problem that one must assume a distance (and mass) when deriving a 
photometric surface gravity.  However, this is only a major issue when 
determining abundances of elements from transitions that are strongly sensitive
to log(g).  The model atmosphere parameters determined here are well--bounded 
by and follow the expected trends of the 10 Gyr isochrones with [Fe/H]=--1.5 
($\alpha$--enhanced) and [Fe/H]=$+$0.5 ([$\alpha$/Fe]=0) shown in 
Figure \ref{f2}.  

We do note that 25/156 (16$\%$) stars in our sample converged to a solution
in which log(g)$>$3--3.5.  The derived higher gravity values suggest some of 
these stars may be foreground lower RGB and subgiants rather than more
evolved bulge RGB stars.  A better measurement of surface gravity, either from 
the addition of more than 2--3 Fe II lines or the inclusion of more sensitive 
atmospheric pressure indicators, would better constrain the true nature of 
these stars.  We do not find any strong systematic differences in 
the derived [X/Fe] ratios between stars of ``low" and ``high" 
gravity\footnote{The [Na/Fe] ratios may be an exception as the high gravity
stars tend to have lower [Na/Fe], on average.}, but it is unclear if the 
similar abundances should have any bearing when interpreting bulge versus 
thin/thick disk composition differences (see Section 4).  However, the high 
gravity stars are also relatively metal--rich $\langle$[Fe/H]$\rangle$=$+$0.09 
($\sigma$=0.25), located preferentially on the blue half of the 
color--magnitude diagrams, and have a relatively small velocity dispersion
($\sigma$=55 km s$^{\rm -1}$ for log(g)$>$3).  These data provide additional
circumstantial evidence that the high gravity stars may be foreground,
though possibly inner disk, contaminators (see also Zoccali et al. 2008; 
their Section 7).

\begin{figure}
\epsscale{1.00}
\plotone{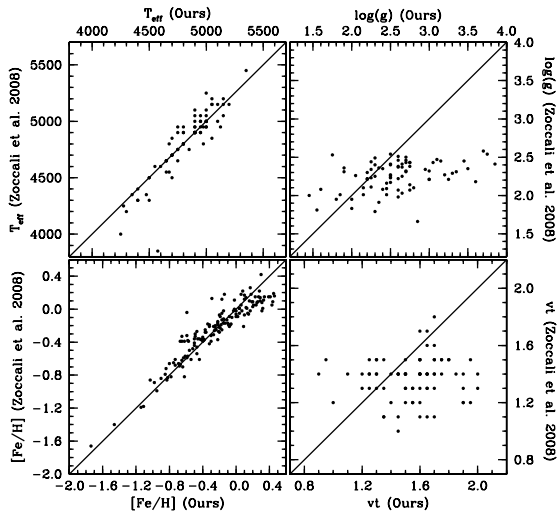}
\caption{Derived model atmosphere parameters are compared between this work
and Zoccali et al. (2008).  Similar to Figure \ref{f2}, the temperature,
gravity, and microturbulence values are only available for the (0,--12) field
in Zoccali et al. (2008).  However, the metallicity panel compares our
results to those in both the (0,--12) (Zoccali et al. 2008) and
($+$5.25,--3.02) (Gonzalez et al. 2011) fields.  In all panels the solid
black line indicates perfect agreement.}
\label{f3}
\end{figure}

In Figure \ref{f3} we compare the derived model atmosphere parameters between
this study, Zoccali et al. (2008), and Gonzalez et al. (2011).  We find good
agreement in the derived T$_{\rm eff}$ values with an average difference
of only 2 K ($\sigma$=98 K).  The dispersion of $\sim$100 K is reasonable given
the different line lists and model atmospheres (but similar technique of 
excitation equilibrium) used.  As mentioned previously, there is some 
discrepancy in log(g), especially for the highest gravity stars, between the 
present work and Zoccali et al. (2008).  For stars with log(g)$<$$+$2.5, the 
average difference in log(g) is 0.01 dex ($\sigma$=0.29 dex), but for stars 
with log(g)$>$$+$2.5 the magnitude of the average gravity difference is 0.64 dex
($\sigma$=0.37 dex).  Comparing the microturbulence values, which may be 
particularly sensitive to line choice and can vary as a function of 
gravity, we find an average difference of 0.18 km s$^{\rm -1}$ ($\sigma$=0.27 
km s$^{\rm -1}$).

\begin{figure}
\epsscale{1.00}
\plotone{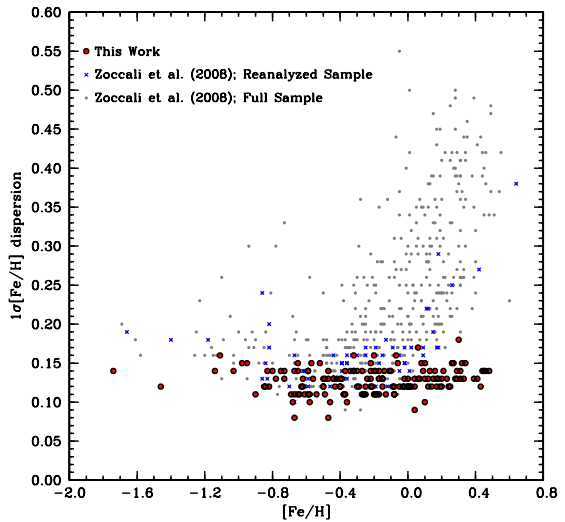}
\caption{The 1$\sigma$ line--to--line dispersion values for [Fe/H] measurements
in this work (filled red circles), the original Zoccali et al. (2008) stars
selected for reanalysis (blue crosses), and the full Zoccali et al. (2008)
sample that includes all four fields (filled gray circles).}
\label{f4}
\end{figure}

When comparing derived [Fe/H] values, we find good agreement
for [Fe/H]$<$$+$0.2 with an average difference of 0.03 dex ($\sigma$=0.13 dex).
However, as is evident in Figure \ref{f3} our derived [Fe/H] values are 
systematically higher on average by 0.18 dex ($\sigma$=0.13 dex), for stars
with [Fe/H]$>$$+$0.2.  The source of this discrepancy may be related to the 
large 1$\sigma$ [Fe/H] uncertainties given in Zoccali et al. (2008) for stars 
with [Fe/H]$\ga$0.  This is illustrated in Figure \ref{f4} where we plot the 
1$\sigma$ [Fe/H] uncertainties between our work and Zoccali et al. (2008) as a 
function of [Fe/H].  Ideally one expects to have measurement errors that are 
not correlated with metallicity, as is the case here.  For the Zoccali et al.
(2008) subsample in common with the present analysis, the line--to--line 
dispersions are comparable only for stars with [Fe/H]$\la$$+$0.2.

\subsection{Equivalent Width Abundance Determinations}

\begin{figure}
\epsscale{0.80}
\plotone{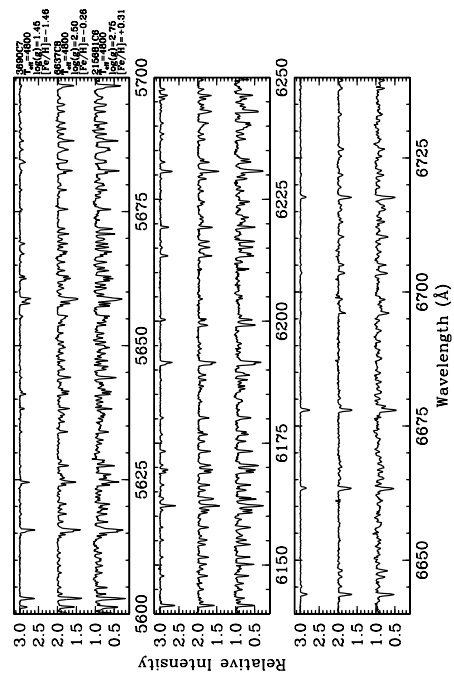}
\caption{Sample spectra are shown to illustrate both data quality and the
change in line strengths and continuum availability for stars of similar
temperature but varying metallicity.}
\label{f5}
\end{figure}

The abundances of Fe I, Fe II, Si I, Ca I, Cr I, and Ni I were determined 
by measuring EWs via an interactive, semi--automatic code developed for this 
project.  The measurement process followed the ``standard" procedure of fitting
single or multiple Gaussian profiles to the spectra for isolated and weakly
blended lines, respectively.  However, the measurement time frame was 
significantly reduced by implementing a simple machine learning algorithm that
kept track of user input on a per--line basis to make an educated first guess
for subsequent measurements in other stars of: the number of profiles to fit, 
profile fitting edges, and the central wavelength, width, and central depth of
all associated nearby features.  While all EW measurements were manually 
inspected, as was mentioned in Section 2 we selected stars from the archival 
data based primarily on S/N considerations in an effort to reduce measurement 
uncertainties.  Sample spectra for stars of similar temperature but different
metallicity are shown in Figure \ref{f5} to illustrate typical data quality 
in the three spectrograph setups.

The line lists for this project were created by visually examining the high
S/N spectra of cool metal--poor and metal--rich giants in the sample, finding 
all isolated and/or weakly blended features for elements of interest, and 
merging the two line list sets.  This was done to ensure that a roughly 
equivalent number of lines could be used in metal--rich and metal--poor stars,
and the manual inspection of each fit enabled us to discard prohibitively
strong and weak lines.  On average, the Fe I, Fe II, Si I, Ca I, Cr I, and 
Ni I abundances were based on the measurement of 70, 2, 8, 6, 6, and 16 lines, 
respectively.  The log(gf) values were set via an inverse abundance analysis 
relative to Arcturus.  We adopted the Arcturus model atmosphere parameters from
Fulbright et al. (2006).  Similarly, for Fe, Si, and Ca we adopted the 
Arcturus abundances from Fulbright et al. (2006), and for Cr and Ni we 
adopted the Arcturus abundances from Ram{\'{\i}}rez \& Allende Prieto (2011).
The final line list, including the adopted Arcturus and derived solar 
abundances (based on measurements of the Hinkle et al. 2000 Arcturus and 
solar atlases), are provided in Table 2.  The derived solar abundances for
Fe, Si, Ca, and Cr agree within $\sim$0.05 dex of the values given in 
Asplund et al. (2009).

The final abundances of Fe I, Fe II, Si I, Ca I, Cr I, and Ni I, determined 
using the \emph{abfind} driver of the LTE line analysis code MOOG (Sneden 1973;
2010 version), are given in Table 3.  Note also that the [Fe/H] values given 
in Table 1 are the average of the [Fe I/H] and [Fe II/H] abundances given in 
Table 3.  However, the average difference in the sense [Fe I/H]--[Fe II/H] is 
$+$0.00 dex with a small dispersion ($\sigma$=0.02 dex).

\subsection{Spectrum Synthesis Abundance Determinations}

For the element abundances derived from transitions involving a small number of 
lines that are affected by significant blends from prevalent spectral features, 
such as molecules and Ca I autoionization, and/or broadened due to isotopes 
and/or hyperfine structure, we used spectrum synthesis rather than EW 
analyses.  For the present work this list includes [O I], Na I, Mg I, Al I,
Co I, and Cu I.  The abundances were determined using the parallelized version 
of the \emph{synth} driver for MOOG (Johnson et al. 2012).  For O, Na, Mg, and
Al, we adopted as a reference point the Arcturus abundances given in 
Fulbright et al. (2006).  However, as described below the reference Arcturus
abundances for Co and Cu are based on measurements using the Kurucz
(1994) and Cunha et al. (2002) hyperfine structure line lists.

The specific reasons for using synthesis are slightly different for each 
element given above.  The 6300.30 \AA\ [O I] line is blended with both a Sc II
feature at 6300.69 \AA\ and a Ni I feature at 6300.33 \AA.  Additionally,
for most stars in this sample the oxygen abundance is sensitive to the 
molecular equilibrium calculations set by the carbon and nitrogen abundances
as well.  Using the CN line list from the Kurucz (1994) database, we 
iteratively solved for the O and C$+$N abundances in each star.  For 
sodium, the 6154.23 Na I line is relatively clean, but the 6160.75 Na I line 
is partially blended with two relatively strong Ca I lines.  The three Mg I 
lines at 6319 \AA\ are strongly affected by a broad Ca I autoionization 
feature, which we set by fitting the slope of the pseudo--continuum from 
$\sim$6316--6318 \AA.  The 6696.02 and 6698.67 \AA\ Al I lines are both 
affected by CN, particularly in cooler and more metal--rich stars.  Therefore, 
as with [O I] we simultaneously fit the Al I doublet and nearby CN features.  
The odd--Z isotope \iso{59}{Co} constitutes almost 100$\%$ of the cobalt 
abundance.  While the 5647.23 and 6117.00 \AA\ Co I lines are relatively weak 
(EW $\la$50 m\AA), we included the hyperfine structure components from the 
Kurucz (1994) line list in our syntheses.  For copper, which is dominated by 
the two odd--Z isotopes \iso{63}{Cu} and \iso{65}{Cu}, we assumed a solar 
system mixture of 69.17$\%$ and 30.83$\%$, respectively.  We adopted the 
hyperfine line list of Cunha et al. (2002) and derive a similar solar abundance
of log $\epsilon$(Cu)=$+$4.04 but a slightly lower Arcturus abundance than 
McWilliam et al. (2013).  Although the 5782.11 \AA\ Cu I line is strong in most
of our stars, the hyperfine broadening helps desaturate the line profile 
to some extent.

\begin{figure}
\epsscale{1.00}
\plotone{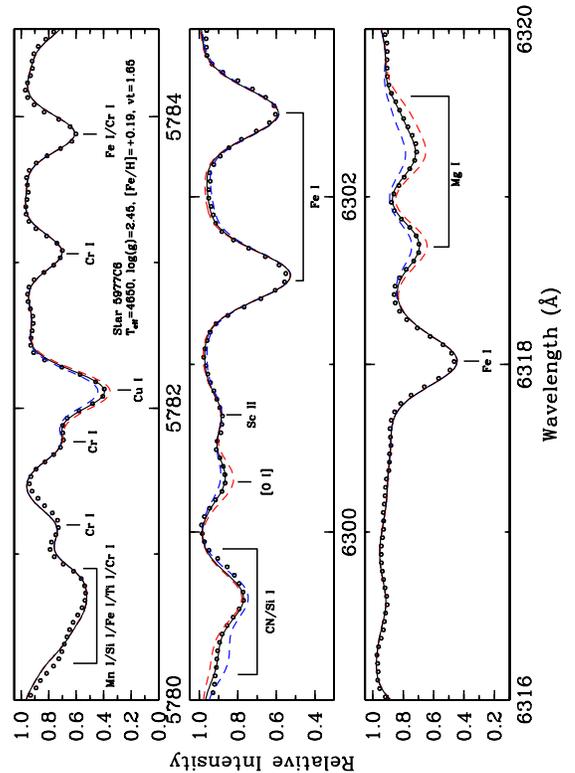}
\caption{Sample spectrum synthesis fits are shown for the Cu I, [O I], and
Mg I features.  In all panels the solid black line indicates the best--fit
value.  The dashed red and blue lines indicate changes to the best--fit
abundance by $\pm$ 0.3 dex, respectively.}
\label{f6}
\end{figure}

In Figure \ref{f6} we show sample syntheses of the O, Mg, and Cu features for a
typical metal--rich spectrum.  We note that the 5782 \AA\ Cu I line is also 
sometimes affected by a nearby diffuse interstellar band (DIB).  The width and 
depth of the DIB feature was found to be highly variable.  The level of 
contamination depends on the relative velocity between the interstellar cloud 
and the individual star and also the reddening value.  Therefore, stars in the 
($+$5.25,--3.02) field, which have an average E(B--V)=0.7, were more strongly 
affected than those in the (0,--12) field, which have an average E(B--V)=0.2 
(Zoccali et al. 2008).  Most of the stars listed in Table 3 that do not have 
a [Cu/Fe] abundance listed were omitted because of obvious contamination by 
the DIB feature.

\subsection{Radial Velocities}

Radial velocities were measured using the XCSAO code (Kurtz \& Mink 1998) for 
each individual exposure of every star and in all three filters.  For 
reference templates we generated synthetic spectra ranging in temperature from 
4250 to 5000 K (250 K steps), log(g) from $+$0.5 to $+$3.5 cgs (0.5 dex steps),
[Fe/H] from --1.5 to $+$0.5 dex (0.5 dex steps), and vt from 1 to 2 
km s$^{\rm -1}$ (0.25 km s$^{\rm -1}$ steps).  Radial velocities were 
determined relative to the nearest template.  We found the average agreement
between exposures to be 0.15 km s$^{\rm -1}$ ($\sigma$=0.13 km s$^{\rm -1}$).
The heliocentric corrections were taken from the headers of the pipeline
reduced files, and the heliocentric radial velocities (RV$_{\rm helio.}$) 
listed in Table 1 represent the average value of all exposures and filters for 
each star.

\begin{figure}
\epsscale{1.0}
\plotone{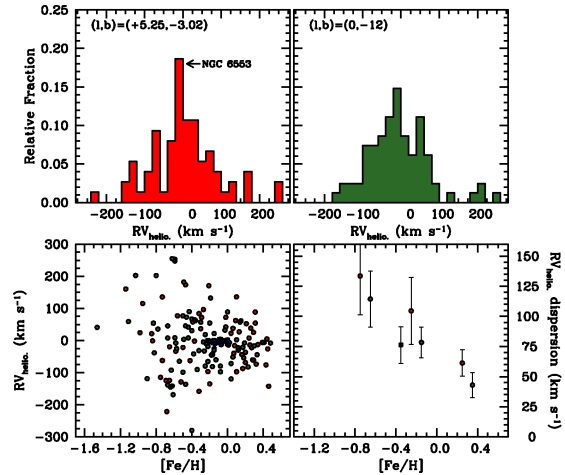}
\caption{\emph{top left:} The red histogram (20 km s$^{\rm -1}$ bins)
illustrates the heliocentric radial velocity distribution function for the
($+$5.25,--3.02) field.  The bulge globular cluster NGC 6553 is labeled.
\emph{top right:} The green histogram (20 km s$^{\rm -1}$ bins) illustrates
the heliocentric radial velocity distribution function for the (0,--12) field.
\emph{bottom left:} Heliocentric radial velocity is plotted as a function of
[Fe/H] for the ($+$5.25,--3.02) and (0,--12) stars, which are shown as filled
red and filled green circles, respectively.  The NGC 6553 stars (filled blue
boxes) are particularly evident in this panel.  \emph{bottom right:} The
heliocentric radial velocity dispersion is plotted as a function of (binned)
[Fe/H], using the same color scheme as the other panels.  For the middle
[Fe/H] bin the blue box and red circle indicate the velocity dispersion with
(blue) and without (red) the NGC 6553 stars included.}
\label{f7}
\end{figure}

\begin{figure}
\epsscale{1.0}
\plotone{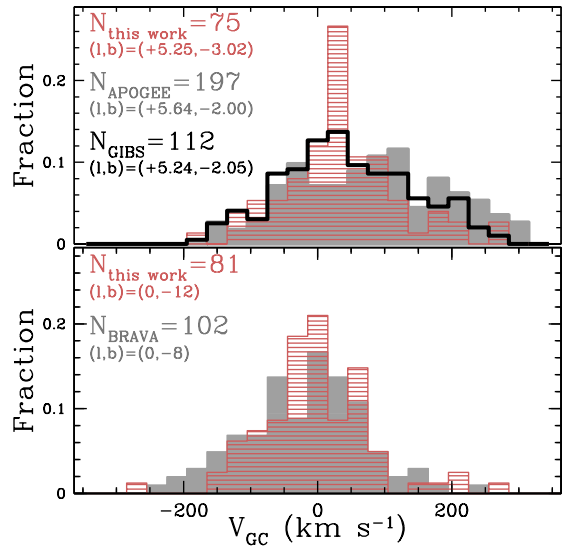}
\caption{\emph{top:} The striped red, solid gray, and black lined histograms
compare the galactocentric radial velocity distributions between the
($+$5.25,--3.02) field analyzed here and nearby fields observed as part of
APOGEE and GIBS, respectively.  The narrow peak near the center of the
distribution is due to NGC 6553.  \emph{bottom:} The striped red and gray
histograms compare the galactocentric radial velocity distributions between
the (0,--12) field analyzed here and the relatively nearby (0,--8) field from
the BRAVA survey.}
\label{f8}
\end{figure}

The kinematic properties of the bulge have been extensively discussed in
dedicated survey papers (e.g., Rich et al. 2007a; Howard et al. 2009; Rangwala 
et al. 2009a; Babusiaux et al. 2010; Kunder et al. 2012; Ness et al. 2013b; 
Nidever et al. 2012; Babusiaux et al. 2014; Zoccali et al. 2014).  Therefore, 
here we seek only to place our results in context with those surveys.  
Figure \ref{f7} shows velocity histograms for both fields, the velocity 
distribution as a function of [Fe/H], and the velocity dispersion as a function
of [Fe/H].  While a detailed comparison between our measured velocities and 
those in Babusiaux et al. (2010) is not possible because their individual 
velocities were not published, for both fields we can compare our average
results with those given in Figure 13 of Zoccali et al. (2008) and Table 3 of 
Babusiaux et al. (2010).  For the ($+$5.25,--3.02) field, ignoring NGC 6553 
stars, we find average velocity and dispersion values of $+$4.55 km 
s$^{\rm -1}$ and 95.51 km s$^{\rm -1}$, respectively.  This compares well with
the Zoccali et al. (2008) average velocity of $+$11 km s$^{\rm -1}$ and 
velocity dispersion of 107 km s$^{\rm -1}$.  Similarly, in the (0,--12) field 
we measured an average heliocentric radial velocity of --8.61 km s$^{\rm -1}$
($\sigma$=85.56 km s$^{\rm -1}$) compared to the Babusiaux et al. (2010) 
values of --14 km s$^{\rm -1}$ ($\sigma$=80 km s$^{\rm -1}$).  Additionally, 
as can be seen in Figure \ref{f8} our galactocentric radial velocity 
(V$_{\rm GC}$) distributions are similar to those of nearby fields from the 
BRAVA, GIBS, and APOGEE surveys.

With the exception of the stars obviously related to NGC 6553, we find in 
agreement with previous bulge studies that, at least away from the Galactic 
plane, the velocity distributions are normal with no evidence for significant 
cold populations (but see also Rangwala et al. 2009b).  This contrasts with 
Nidever et al. (2012) and Babusiaux et al. (2014), which find kinematically 
cold populations with V$_{\rm GC}$$\sim$$+$200 km s$^{\rm -1}$.  However, their
fields are significantly closer to the Galactic plane than those analyzed here.
We do note however that these high velocity populations are also not found in 
the BRAVA, ARGOS, nor GIBS analyses, nor is there yet a satisfactory 
theoretical explanation for their origin (e.g., Li et al. 2014).

For the (0,--12) field we observe the same trend of a decrease in velocity
dispersion with increasing [Fe/H] found by Babusiaux et al. (2010).  However,
while Babusiaux et al. (2010) find an increase in velocity dispersion with
increasing [Fe/H] in the ($+$1.1,--4) field of Baade's window, our off--axis
but similar Galactic latitude field at ($+$5.25,--3.02) still exhibits a trend
of decreasing velocity dispersion with increasing [Fe/H].  This further 
contrasts with recent fields observed close to the plane by Babusiaux et al.
(2014; see their Figure 16) that also show a possible increase in velocity 
dispersion with increasing [Fe/H].\footnote{However, when taking into account
the error bars in velocity dispersion for the points in Figure 16 of 
Babusiaux et al. 2014, their trend is mostly flat.}  Our result is more similar
to studies of outer bulge fields that find a consistent decrease in velocity 
dispersion with increasing [Fe/H] (e.g., Johnson et al. 2011; Uttenthaler et 
al. 2012; Johnson et al. 2013; Ness et al. 2013b).  There is weak evidence in 
Figure \ref{f7} that the trend in velocity dispersion and [Fe/H] may be more 
shallow for the ($+$5.25,--3.02) field compared to the (0,--12) field.  Note 
that the inclusion (or not) of NGC 6553 stars significantly affects the 
velocity dispersion of the [Fe/H] bin in which the cluster resides.

\subsection{Identifying NGC 6553 Members}

Members of the globular cluster NGC 6553 in the ($+$5.25,--3.02) field are 
best identified in the velocity--metallicity diagram in Figure \ref{f7}.
The likely members (12 stars total) are clustered near [Fe/H]$\approx$--0.10 
and RV$_{\rm helio.}$$\approx$0 km s$^{\rm -1}$.  Literature measurements of 
the cluster's average [Fe/H] value vary considerably, with estimates that 
include: --0.55 (Barbuy et al. 1999), --0.16 (Cohen et al. 1999), --0.7 (Coelho
et al. 2001), --0.3 (Origlia et al. 2002), --0.2 (Mel{\'e}ndez et al. 2003), and
--0.2 (Alves--Brito et al. 2006).  However, we find in agreement with the 
most recent estimates that $\langle$[Fe/H]$\rangle$=--0.11 ($\sigma$=0.07).
While the cluster is slightly iron--deficient relative to the Sun, the moderate
enhancements of the cluster's [$\alpha$/Fe] ratio (see Section 4.1) gives it 
an overall metallicity that is roughly solar.  NGC 6553 is
therefore one of the most metal--rich globular clusters in the Galaxy.

We find similar agreement with literature values for the cluster's radial 
velocity, with $\langle$RV$_{\rm helio.}$$\rangle$=--2.03 km s$^{\rm -1}$ 
($\sigma$=4.85 km s$^{\rm -1}$).  This is compared with recent values of:
--1 km s$^{\rm -1}$ (Coelho et al. 2001), $+$1.6 km s$^{\rm -1}$ (Mel{\'e}ndez 
et al. 2003), and --1.86 km s$^{\rm -1}$ (Alves--Brito et al. 2006).  Finally,
we note that the stars identified in Table 1 as possible cluster members 
have an average, projected radial distance from the cluster center of 
about 6$\arcmin$ ($\sigma$=5$\arcmin$).  We have adopted a more lenient 
radial distance discriminator than the 2$\arcmin$ limit used by Zoccali et al. 
(2008) and Gonzalez et al. (2011), and instead rely more on the [Fe/H] and 
velocity measurements to identify possible cluster members.

\subsection{Abundance Ratio Comparisons with Previous Work}

\begin{figure}
\epsscale{1.0}
\plotone{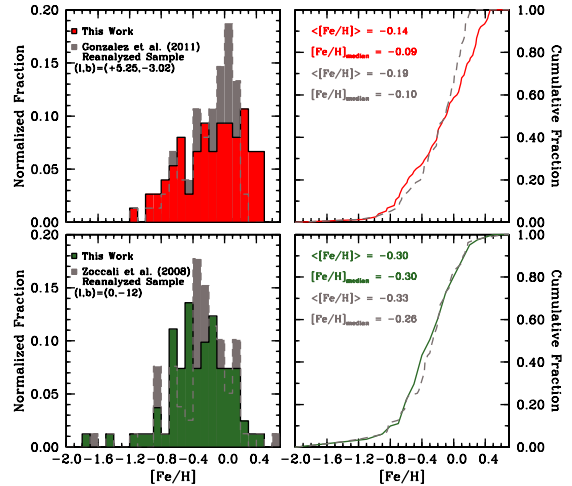}
\caption{\emph{top left:} The red and gray histograms (0.1 dex bins) illustrate
the derived metallicity distribution functions for the ($+$5.25,--3.02) field
in this work and Gonzalez et al. (2011), respectively.  \emph{top right:} The
solid red and dashed gray lines illustrate the cumulative distribution
functions for this work and Gonzalez et al. (2011), respectively.  \emph{bottom
left:} The green and gray histograms (0.1 dex bins) illustrate the derived
metallicity distribution functions for the (0,--12) field in this work and
Zoccali et al. (2008), respectively.  \emph{bottom right:}  The solid green
and dashed gray lines illustrate the cumulative distribution functions for
this work and Zoccali et al. (2008), respectively.}
\label{f9}
\end{figure}

\begin{figure}
\epsscale{1.0}
\plotone{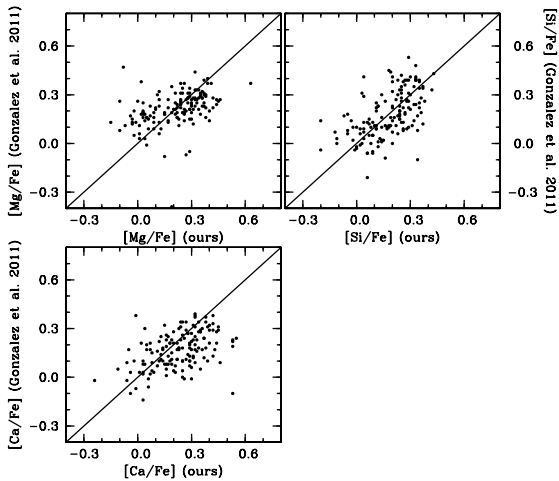}
\caption{A comparison between the [Mg/Fe], [Si/Fe], and [Ca/Fe] abundances
derived here and in Gonzalez et al. (2011).  The solid black line in each
panel indicates perfect agreement.}
\label{f10}
\end{figure}

As noted previously, Zoccali et al. (2008) and Gonzalez et al. (2011) presented
[Fe/H], [Si/Fe], [Ca/Fe], and [Ti/Fe] abundances based on the same GIRAFFE
data utilized here.  Therefore, in Figures \ref{f9}--\ref{f10} we compare 
our results with theirs for stars and elements in common.  While a quantitative
comparison of the individual [Fe/H] values is given in Section 3.1 (see also
Figure \ref{f3}), in Figure \ref{f9} we compare the general shapes and 
bulk properties of the metallicity distribution functions.  For the 
($+$5.25,--3.02) field the average and median [Fe/H] ratios are similar, but
the distribution from the present work is somewhat broader and extends to 
higher [Fe/H].  In contrast, there are no significant differences in the 
[Fe/H] distribution functions between the present work and the same stars
from Zoccali et al. (2008), for the (0,--12) field.  We also reconfirm one of 
the primary conclusions of Zoccali et al. (2008) that interior bulge fields 
have a higher average metallicity than outer bulge fields.  Finally, we note 
that the distribution functions shown in Figure \ref{f9} do not provide 
strong evidence supporting the existence of multiple, discreet populations,
as has been suggested in some studies (Bensby et al. 2011; Hill et al. 2011; 
Bensby et al. 2013; Ness et al. 2013a).  However, the number of stars per 
field presented here is $<$100.

In Figure \ref{f10} we compare our derived [Mg/Fe], [Si/Fe], and [Ca/Fe] ratios
to those given in Gonzalez et al. (2011).  The average differences 
between the present work and that of Gonzalez et al. (2011) are 
$\Delta$[Mg/Fe]=$+$0.00 ($\sigma$=0.14), $\Delta$[Si/Fe]=$+$0.00 
($\sigma$=0.13), and $\Delta$[Ca/Fe]=$-$0.06 ($\sigma$=0.14).  The relatively
consistent star--to--star scatter of $\sim$0.14 dex is a reasonable estimate 
of the attainable precision between the two studies, which derive 
$\alpha$--element abundances from different techniques (synthesis in Gonzalez
et al. 2011 and EW measurements here).  We note that the $\alpha$--elements
oxygen (measured here) and titanium (measured in Gonzalez et al. 2011) were 
not both measured in each study.\footnote{Gonzalez et al. (2011) did not
derive an oxygen abundance from the 6300 \AA\ [O I] feature because of 
concerns regarding measurement accuracy.  We chose not to include Ti abundances
because of discrepant nucleosynthesis predictions for this element in 
comparison to observations (e.g., see Kobayashi et al. 2011; their Figure 14).}

\subsection{Abundance Uncertainty Estimates}

We investigated the sensitivity of derived abundances for 
each element in every star by taking the abundances given in Table 3, 
determining theoretical EWs using the line list in Table 2, and then 
varying the model atmosphere parameters T$_{\rm eff}$, log(g), [Fe/H], and 
vt individually while holding the other parameters fixed.  We selected 
parameter changes of 100 K in T$_{\rm eff}$, 0.30 dex in log(g), 0.15 dex in 
[M/H], and 0.30 km s$^{\rm -1}$ in vt, which are reasonable when comparing our 
derived parameters with those of the independent analysis by Zoccali et al. 
(2008; see also Section 3.1).  The total uncertainty for each element ratio in 
each star resulting from this exercise is provided in Table 4.

In general, most elements are not affected by changes in T$_{\rm eff}$ of 100 K
at more than the 0.1 dex level.  However, the two species presented here that 
reside in their dominant ionization states ([O I] and Fe II) are strongly 
affected by changes in surface gravity.  For a change in log(g) of 0.3 dex, the 
log $\epsilon$(O) and log $\epsilon$(Fe II) abundances can change by 
$\sim$0.1--0.3 dex, but these effects are mitigated when the [O I/H] abundance
is normalized with [Fe II/H].  These two species are also more strongly 
affected by changes in the model metallicity, and the larger [Fe II/H] 
measurement and sensitivity uncertainties are a contributing factor to the 
increased dispersion in the [O/Fe] ratios compared to other $\alpha$--elements 
(e.g., [Mg/Fe]).  As expected, microturbulence sensitivity is correlated with a 
star's overall metallicity (i.e., line strength).  Among the transitions under 
consideration here, in metal--rich stars those of Na, Ca, and Cu typically 
have the strongest lines and are thus more strongly affected by the 
microturbulence uncertainty.

In Table 5 we also provide the 1$\sigma$ line--to--line dispersion values for
all species measured here.  These values should be mostly representative of the
combined measurement error that includes effects such as: continuum placement, 
line deblending, synthesis fits via visual inspection, log(gf) uncertainties,
and model atmosphere deficiencies.  Typical line--to--line dispersion values
are $\sim$0.08 dex.  The measurement error of Cu may be underestimated 
because of the line's large EW, non--negligible blending (see Figure \ref{f6}),
and possible contamination with a nearby DIBS feature.  A more reasonable 
measurement uncertainty for Cu is, in most cases, $\sim$0.15--0.20 dex.

\section{RESULTS AND DISCUSSION}

\subsection{The $\alpha$ Elements Oxygen, Magnesium, Silicon, and Calcium}

The $\alpha$--elements have been the primary focus of detailed composition work
in the Galactic bulge.  To first order there is agreement among the various 
studies that: (1) the [$\alpha$/Fe] ratios are enhanced by $\sim$$+$0.3 dex at 
[Fe/H]$\la$--0.3, (2) for stars with [Fe/H]$\ga$$+$0.3 there is a mostly 
monotonic decline in [$\alpha$/Fe] with increasing [Fe/H], (3) the bulge 
and thick disk may share similar chemistry over a wide range in metallicity,
and (4) there are no significant variations in the [$\alpha$/Fe] trends 
between different bulge sight lines (McWilliam \& Rich 1994; Cunha \& Smith 
2006; Zoccali et al. 2006; Fulbright et al. 2007; Lecureur et al. 2007;
Mel{\'e}ndez et al. 2008; Alves--Brito et al. 2010; Bensby et al. 2010a;
Ryde et al. 2010; Bensby et al. 2011; Gonzalez et al. 2011; Hill et al. 2011;
Johnson et al. 2011; Uttenthaler et al. 2012; Bensby et al. 2013; 
Garc{\'{\i}}a P{\'e}rez et al. et al. 2013; Johnson et al. 2013; Ness et al.
2013a; J{\"o}nsson et al. 2014).  Additionally, there is evidence that the 
[O/Mg] ratio declines with increasing metallicity (Fulbright et al. 2007; 
Lecureur et al. 2007; McWilliam et al. 2008; Alves--Brito et al. 2010).

\begin{figure}
\epsscale{1.0}
\plotone{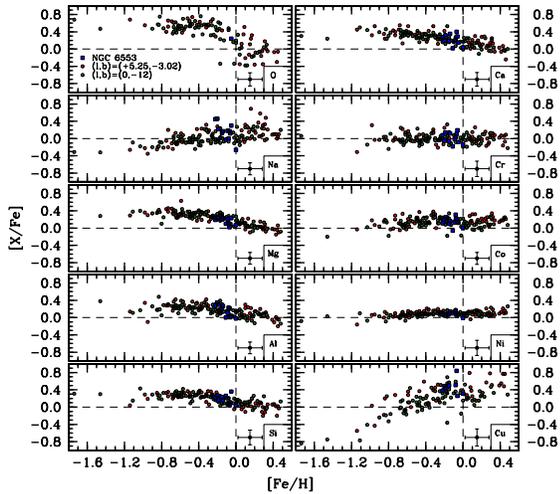}
\caption{[X/Fe] abundance patterns plotted as a function of [Fe/H] for all
elements analyzed.  The filled red circles, filled green circles, and filled
blue boxes differentiate stars belonging to the ($+$5.25,--3.02), (0,--12),
and NGC 6553 populations.  Note that the scale of the ordinate is identical in
all panels.}
\label{f11}
\end{figure}

\begin{figure}
\epsscale{1.0}
\plotone{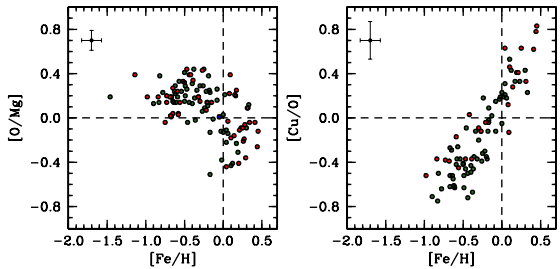}
\caption{The [O/Mg] and [Cu/O] ratios are plotted as a function of [Fe/H].
The filled red circles, filled green circles, and filled blue boxes
differentiate stars belonging to the ($+$5.25,--3.02), (0,--12), and NGC 6553
populations.}
\label{f12}
\end{figure}

The new data presented here, and summarized in Figure \ref{f11}, reinforce many
observations from the previous studies mentioned above.  In particular, we find
that for [Fe/H]$<$--0.3 all of the [$\alpha$/Fe] ratios are enhanced and 
exhibit minimal star--to--star scatter with $\langle$[O/Fe]$\rangle$=$+$0.54 
($\sigma$=0.10), $\langle$[Mg/Fe]$\rangle$=$+$0.33 ($\sigma$=0.08), 
$\langle$[Si/Fe]$\rangle$=$+$0.28 ($\sigma$=0.07), and 
$\langle$[Ca/Fe]$\rangle$=$+$0.34 ($\sigma$=0.09).  For stars with 
[Fe/H]$>$--0.3, we find that the [$\alpha$/Fe] ratios decrease with increasing
[Fe/H].  However, Figure \ref{f11} illustrates the disparate trends for 
individual elements and highlights the information loss that can occur when
averaging abundance ratios for multiple $\alpha$--elements.  The [O/Fe] ratios 
are higher by $\sim$0.2 dex in metal--poor stars than those of other 
$\alpha$--elements, but this trend reverses for stars with [Fe/H]$\ga$0 where 
[O/Fe] is, on average, lower by $\sim$0.2 dex.  While both O and Mg are 
significant products of hydrostatic burning in massive stars (e.g., Woosley \& 
Weaver 1995), the [Mg/Fe] trend exhibits a more shallow decline with 
increasing [Fe/H] than [O/Fe].  This is most clearly seen in Figure \ref{f12},
which shows a sharply declining [O/Mg] ratio at [Fe/H]$\ga$--0.1.  Although
massive star production of Si and Ca involves both hydrostatic and explosive 
burning (e.g., Woosley \& Weaver 1995), the [Si/Fe] and [Ca/Fe] trends are more
similar to [Mg/Fe] than [O/Fe].  Given the disparate trend of [O/Fe] compared
to other $\alpha$--elements, and the low production of most $\alpha$--elements
relative to Fe in Type Ia supernovae (SNe; e.g., Nomoto et al. 1997), we 
conclude in agreement with past work (e.g., McWilliam et al. 2008) that the 
strong decline in [O/Fe] at [Fe/H]$\ga$--0.3 is likely a result of metallicity 
dependent yields in massive stars\footnote{We note that for standard stellar
evolution models it is often difficult to produce a strong change in the 
[O/Mg] yield (e.g., Kobayashi et al. 2006; their Figure 9).  Producing models
with a declining [O/Mg] yield likely requires the inclusion of additional
physics, such as mass loss from stellar winds or a process to cause a change
in the [C/O] ratio.}.  The influence of metallicity dependent yields on the 
bulge composition profile, especially from Wolf--Rayet stars, is further 
supported by fluorine measurements (Cunha et al. 2008; J{\"o}nsson et al. 
2014), but it remains to be seen if this scenario can be reconciled with the 
observed carbon and nitrogen trends (Ryde et al. 2010; but see also Cescutti 
et al. 2009).

\begin{figure}
\epsscale{1.0}
\plotone{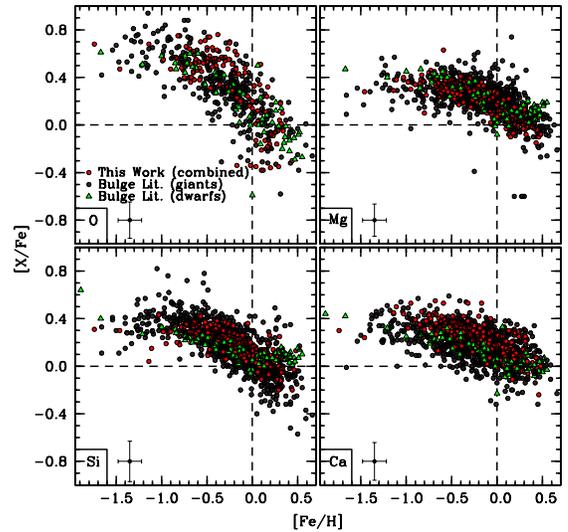}
\caption{[X/Fe] ratios for the $\alpha$--elements O, Mg, Si, and Ca are plotted
as a function of [Fe/H].  The filled red circles indicate abundances measured
for this work (combining both fields and NGC 6553), the filled dark gray
circles are abundances in bulge RGB and red clump stars from the literature,
and the filled green triangles are abundances from bulge microlensed dwarfs
(Bensby et al. 2013).  The RGB and clump data are from: McWilliam \& Rich
(1994), Rich et al. (2005), Fulbright et al. (2007), Lecureur et al. (2007),
Rich et al. (2007b), Mel{\'e}ndez et al.(2008), Alves--Brito et al. (2010),
Ryde et al. (2010), Gonzalez et al. (2011), Hill et al. (2011), Johnson et
al. (2011), Rich et al. (2012), Garc{\'{\i}}a P{\'e}rez et al. (2013), and
Johnson et al. (2013a,b).}
\label{f13}
\end{figure}

When comparing the individual [$\alpha$/Fe] trends between the two fields 
analyzed here, Figure \ref{f11} shows no significant variations.  Similarly,
in Figure \ref{f13} we combine our two fields and compare with literature 
giant and dwarf [$\alpha$/Fe] data.  A comparison between the present work
and literature giant trends, which span a variety of bulge sight lines, leads 
us to find in agreement with Johnson et al. (2011; 2013) and Gonzalez et al.
(2011) that no significant field--to--field [$\alpha$/Fe] variations exist over
a broad region of the bulge.  The microlensed dwarf data exhibit the same 
qualitative and quantitative distributions for [O/Fe] and [Mg/Fe], at least for
[Fe/H]$\la$0, as the giant data, but there may be small systematic offsets
with [Si/Fe] and [Ca/Fe].  In particular, the dwarf abundances are $\sim$0.1 
dex lower for a given [Fe/H], when considering [Fe/H]$\la$0.  At super--solar 
[Fe/H] values, the dwarf and giant data are in excellent agreement for [O/Fe], 
but the leveling--off or slight increase in [Mg/Fe], [Si/Fe], and [Ca/Fe] seems
to be unique to the dwarf measurements.  Unfortunately, the source of this 
discrepancy is not clear and may be related to analysis differences 
between dwarfs and giants.

\begin{figure}
\epsscale{1.0}
\plotone{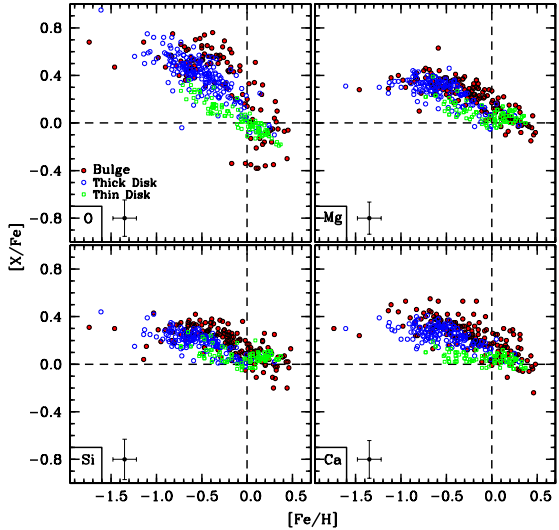}
\caption{A comparison of the O, Mg, Si, and Ca abundances for the bulge
stars measured here (filled red circles) with those of the thick disk (open
blue circles) and thin disk (open green boxes).  The disk data are from: Bensby
et al. (2003; 2005) and Reddy et al. (2006).}
\label{f14}
\end{figure}

In Figure \ref{f14} we compare the [O/Fe], [Mg/Fe], [Si/Fe], and [Ca/Fe] 
abundances between the bulge, thick disk, and thin disk.  For stars with 
[Fe/H]$\la$--0.5, the bulge and thick disk stars exhibit similar abundance
patterns for all four element ratios.  However, we note that on average the 
bulge stars have [O/Fe] and [Mg/Fe] ratios that are slightly enhanced by 
$\sim$0.03 dex and [Si/Fe] and [Ca/Fe] ratios that are enhanced by $\sim$0.05 
dex compared to similar metallicity thick disk stars.  In contrast, the most
metal--poor thin disk stars exhibit significantly lower [X/Fe] ratios for
all of the $\alpha$--elements measured here.  The bulge and thin
disk stars with [Fe/H]$\ga$0 are not strikingly different, but the 
star--to--star scatter, especially for [O/Fe], is significantly larger for 
the bulge giants.  For the intermediate range of [Fe/H]$\sim$--0.5 to 0, the 
bulge stars still exhibit significantly larger [$\alpha$/Fe] ratios than the 
thin disk and may remain enhanced to a higher [Fe/H] value than the thick disk.

The chemical similarities between especially the metal--poor bulge and thick 
disk found here have been documented in previous work (Mel{\'e}ndez et 
al. 2008; Alves--Brito et al. 2010; Bensby et al. 2010a; Ryde et al. 2010; 
Bensby et al. 2011; Gonzalez et al. 2011; Hill et al. 2011; Johnson et al. 
2011, 2013).  The apparent homogeneity between the most metal--poor bulge and 
thick disk stars lends credibility to the idea that the Galactic bulge formed 
\emph{in situ} with the disk.  However, there is not universal agreement in the
literature that the metal--poor bulge and disk trends are identical.  In 
particular, earlier work by Zoccali et al. (2006), Fulbright et al. (2007), 
and Lecureur et al. (2007) found that the bulge stars exhibited both larger
[$\alpha$/Fe] ratios and remained enhanced to higher [Fe/H] values than the
local thick disk\footnote{A comparison with the inner disk would be more 
appropriate; however, we note that Bensby et al. (2010b) do not find any 
significant chemical differences between local and inner disk stars.}, which
implies a more rapid formation timescale for the bulge.  In contrast, purely
differential analyses between thick disk and bulge giants (Mel{\'e}ndez et al. 
2008; Alves--Brito et al. 2010; Gonzalez et al. 2011) find nearly identical 
[$\alpha$/Fe] versus [Fe/H] trends, at least for [Fe/H]$\la$--0.3.  However, 
Bensby et al. (2013) noted in a similarly differential comparison of local 
thick disk dwarfs and bulge microlensed dwarfs that the inflection point at 
which [$\alpha$/Fe] declines may be 0.1--0.2 dex higher ([Fe/H]$\approx$--0.3 
to --0.2) in the bulge.  While we compare bulge giants and thick disk dwarfs in 
Figure \ref{f14}, our results are in agreement with Bensby et al. (2013).
In particular, we find that the [Mg/Fe], [Si/Fe], and perhaps [O/Fe] ratios
remain enhanced to a higher [Fe/H] value than those of the local thick 
disk\footnote{If we instead compare the [$\alpha$/Fe] ratios between bulge
giants here and thick disk giants from Alves--Brito et al. (2010), we reach
a similar conclusion.  Both data sets exhibit similar abundance trends for
[O/Fe] and [Mg/Fe], but [Si/Fe] and [Ca/Fe] remain enhanced at higher [Fe/H]
in the bulge giants.}.  

Finally, we note that combining the present data set with those available in
the literature (e.g., see Figure \ref{f13}) totals $\sim$10$^{\rm 3}$ bulge
stars that have had [$\alpha$/Fe] measurements made from high resolution 
spectroscopy.  Despite the large sample size, there is a paucity of 
stars with [$\alpha$/Fe] ratios that deviate significantly from the bulk 
trend.  In agreement with work suggesting the Galactic bulge did not form 
predominantly from a build--up of merger events (e.g., Shen et al. 2010), we 
can effectively rule out significant contributions from the infall of objects 
with chemistry similar to those of many present--day dwarf galaxies (i.e., low 
[$\alpha$/Fe]; e.g., see Venn et al. 2004 and references therein).  
Additionally, as can be seen in Figure \ref{f11} (see also Gonzalez et al. 
2011) the [X/Fe] abundance ratios of individual $\alpha$--elements for NGC 6553
stars are nearly identical to those of bulge field stars with similar [Fe/H].  
Specifically, the average [X/Fe] values for NGC 6553 are:
$\langle$[O/Fe]$\rangle$=$+$0.24 (one star), $\langle$[Mg/Fe]$\rangle$=$+$0.16
($\sigma$=0.08), $\langle$[Si/Fe]$\rangle$=$+$0.17 ($\sigma$=0.10), and
$\langle$[Ca/Fe]$\rangle$=$+$0.22 ($\sigma$=0.12), which compare well with
the average abundances for nearby bulge field stars in the range [Fe/H]=--0.20 
to $+$0.00: $\langle$[O/Fe]$\rangle$=$+$0.24 ($\sigma$=0.29), 
$\langle$[Mg/Fe]$\rangle$=$+$0.25 ($\sigma$=0.09),
$\langle$[Si/Fe]$\rangle$=$+$0.15 ($\sigma$=0.08), and
$\langle$[Ca/Fe]$\rangle$=$+$0.19 ($\sigma$=0.13).  These values are in good 
agreement with past work that finds the cluster to be moderately 
$\alpha$--enhanced (Barbuy et al. 1999; Cohen et al. 1999; Coelho et al. 2001;
Origlia et al. 2002; Mel{\'e}ndez et al. 2003; Alves--Brito et al. 2006).  The 
similar [$\alpha$/Fe] abundances between the cluster and field stars suggests 
that NGC 6553 likely formed \emph{in situ} with the bulge field population and 
is not a captured cluster.

\subsection{The Light, Odd--Z Elements Sodium and Aluminum}

In a similar fashion to the $\alpha$--elements, the light, odd--Z elements
Na and Al provide clues of the processes that dominated the chemical 
enrichment of a stellar population.  Furthermore, these elements are useful
for ``chemical tagging" analyses, and both the [Na/Fe] and [Al/Fe] ratios can
vary significantly between stellar populations that have otherwise identical
[$\alpha$/Fe] and [Fe/H] values.  The large ($\ga$0.5 dex) star--to--star 
[Na/Fe] and [Al/Fe] abundance variations present in metal--poor globular 
cluster but not halo/disk stars of the same metallicity are perhaps the most
well--known example of this phenomenon (e.g., see reviews by Gratton et al. 
2004; 2012 and references therein).  While the production of Na and Al is 
dominated by hydrostatic helium, carbon, and neon burning in massive stars,
the final yields are expected to grow significantly with increasing progenitor
mass and metallicity (e.g., Woosley \& Weaver 1995; Kobayashi et al. 2006; 
2011).  Intermediate mass ($\sim$4--8 M$_{\rm \odot}$) asymptotic giant branch
(AGB) stars and the hydrogen--rich envelopes of massive stars can also produce
significant amounts of Na and Al via the NeNa and MgAl proton--capture cycles
(e.g., Decressin et al. 2007; de Mink et al. 2009; Ventura \& D'Antona 2009; 
Karakas 2010).  Since Na and Al are thought to result from similar production 
mechanisms, we expect their abundance patterns to reflect a comparable 
morphology.

While the bulge abundance patterns of [Na/Fe] and [Al/Fe] have not been 
investigated to the extent of the $\alpha$--elements, the combined literature
sample now totals of order a few hundred stars.  Interestingly, the agreement
between studies regarding the [Na/Fe] and [Al/Fe] trends is worse than for 
the $\alpha$--elements.  While all high--resolution analyses (McWilliam \& Rich
1994; Fulbright et al. 2007; Lecureur et al. 2007; Alves--Brito et al. 2010;
Bensby et al. 2010a, 2011; Johnson et al. 2012; Bensby et al. 2013) tend to
agree that the average [Na/Fe] ratio rises with increasing metallicity,
significant scatter is present at [Fe/H]$\la$--1 and [Fe/H]$\ga$0.  Similarly,
there is general agreement that [Al/Fe] is enhanced in bulge stars at 
[Fe/H]$\la$--0.3.  However, some studies find that [Al/Fe] remains enhanced
at super--solar metallicities (McWilliam \& Rich 1994; Fulbright et al. 2007; 
Lecureur et al. 2007; Alves--Brito et al. 2010) while others find a decline in
[Al/Fe], similar to [$\alpha$/Fe] (Bensby et al. 2011, 2011; Johnson et al.
2012; Bensby et al. 2013).  Additionally, there is general agreement that 
the [Na/Fe] and [Al/Fe] trends as a function of [Fe/H] are similar between
the bulge and disk over a broad metallicity range, but differences could be
present at the metal--poor and metal--rich ends of the bulge distribution.
It is also not yet clear if any significant [Na/Fe] and [Al/Fe] abundance
differences exist between different bulge sight lines.

\begin{figure}
\epsscale{1.0}
\plotone{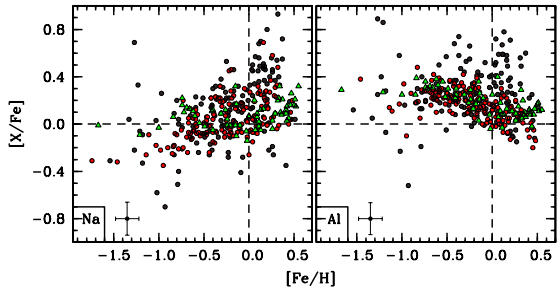}
\caption{A comparison plot of [Na/Fe] and [Al/Fe] ratios between the bulge
RGB stars measured here (filled red circles), RGB and clump stars available
in the literature (filled dark gray circles), and bulge microlensed dwarfs
(filled green triangles).  Additional dwarf and giant literature
data are from: Johnson et al. (2007; 2008), Cohen et al. (2008; 2009),
Epstein et al. (2010), and Johnson et al. (2012), in addition to those
referenced in Figure \ref{f13}.}
\label{f15}
\end{figure}

Figure \ref{f11} shows our derived [Na/Fe] and [Al/Fe] abundances as a function
of [Fe/H] for both fields and the possible NGC 6553 stars, and in 
Figure \ref{f15} we compare our results with those from previous work.  
For Na we find general agreement with literature values such that the average
[Na/Fe] ratio rises with increasing [Fe/H].  However, we find only a small
number of metal--rich stars with [Na/Fe]$>$$+$0.4 and do not reproduce the very 
large [Na/Fe] ratios of Lecureur et al. (2007).  Additionally, we do not find
significant evidence supporting large [Na/Fe] variations between the two
bulge sight lines probed here.  At [Fe/H]$\ga$--0.5, the mean [Na/Fe] trend
and star--to--star dispersion for our measured RGB stars is in good agreement
with those of the microlensed bulge dwarfs (e.g., Bensby et al. 2013).  

\begin{figure}
\epsscale{1.0}
\plotone{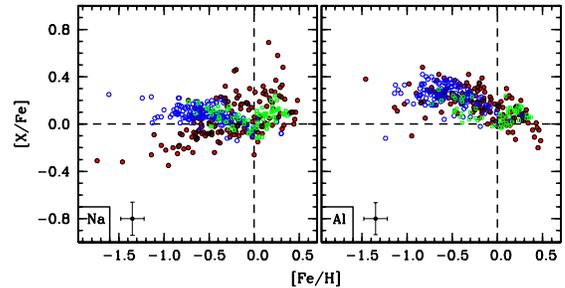}
\caption{A plot of [Na/Fe] and [Al/Fe] ratios as a function of [Fe/H] for
the bulge stars measured here (filled red circles), thick disk stars (open blue
circles), and thin disk stars (open green boxes).  The literature data are from
the sources referenced in Figure \ref{f14}.}
\label{f16}
\end{figure}

The primary discrepancy between our work and some of the literature values
occurs for stars with [Fe/H]$\la$--0.7, with the present work and Johnson et 
al. (2012) finding that the average Na trend decreases from [Na/Fe]$\sim$0 at 
[Fe/H]=--0.5 to [Na/Fe]=--0.3 at [Fe/H]=--1.7.  It is not immediately clear
if the discrepancy, especially between the bulge RGB and dwarf data, is real
or caused by analysis differences (e.g., NLTE, 3D, or spherical/plane--parallel
effects between dwarfs and giants).  The inclusion of NLTE corrections would 
minimize the differences at low metallicity between bulge RGB and dwarf stars, 
and also between bulge RGB and metal--poor thick disk dwarfs (see 
Figure \ref{f16}), if the largely positive Na corrections for RGB stars from 
Gratton et al. (1999) were applied.  However, more recent NLTE calculations 
(e.g., Lind et al. 2011) instead find that the sign of the Na correction is 
negative for the lines and atmospheric parameters used here.  Similarly, the
NLTE corrections for log $\epsilon$(Fe I) appear to be positive (e.g., 
Lind et al. 2012; Bergemann et al. 2012) for most stars in our 
sample\footnote{The NLTE corrections for a subset of ``typical" stars and lines
were calculated using the INSPECT website: http://www.inspect-stars.net.}, 
which would decrease the [Na/Fe] ratios.  Further insights into this problem
may be gained as more extensive NLTE calculations and 3D model atmosphere 
grids and codes become available.

When comparing the [Na/Fe] and [Al/Fe] trends in Figure \ref{f11}, it is
immediately clear that the two elements exhibit discrepant trends.  While
[Na/Fe] gradually rises with increasing [Fe/H], the [Al/Fe] trend is nearly
indistinguishable from that of most $\alpha$--elements.  In particular, we 
find in agreement with Bensby et al. (2010a; 2011; 2013) and Johnson et al. 
(2012) that [Al/Fe]$\sim$$+$0.3 in bulge stars until [Fe/H]$\sim$--0.3 and 
then steadily declines at higher [Fe/H].  As mentioned previously, the 
decline in [Al/Fe] with increasing metallicity contrasts with other literature
results that find [Al/Fe] remains enhanced even at [Fe/H]=$+$0.5 (Fulbright
et al. 2007; Lecureur et al. 2007; Alves--Brito et al. 2010).  The data quality
among the various studies is comparable, and it is not clear why the derived
[Al/Fe] trends are in disagreement at high metallicity.  We do note however 
that for cool, high metallicity stars the 6696 and especially 6698 \AA\ Al I 
lines, as well as the continuum placement, can be affected by CN blending.  

The discrepant [Na/Fe] and [Al/Fe] trends as a function of [Fe/H] are not 
limited to the bulge and may also be present in the disk, as can be seen in
Figure \ref{f16}.  Despite nucleosynthesis models predicting similar production
of Na and Al in massive stars (e.g., Woosley \& Weaver 1995), Figure \ref{f16}
shows that, at least in the metallicity range probed here, Al is over--produced
relative to Na in both bulge and disk stars for [Fe/H]$\la$--0.3.  The 
increased production of Na relative to Al in metal--rich stars, and especially 
in the bulge, suggests that metallicity dependent yields from massive stars 
vary more strongly for Na than Al.  Contributions from intermediate mass AGB 
stars may also help explain the Na and Al trends since the AGB [Na/Fe] yields 
tend to increase at higher [Fe/H] while those of [Al/Fe] decline (e.g., Ventra 
\& D'Antona 2009).  Interestingly, we find that, unlike the case for [Na/Fe], 
the [Al/Fe] ratios are nearly indistinguishable between the bulge and thick 
disk at [Fe/H]$<$0.  Similarly, the [Al/Fe] ratios for bulge stars are 
identical to those in the thin disk at [Fe/H]$>$0.

\begin{figure}
\epsscale{1.0}
\plotone{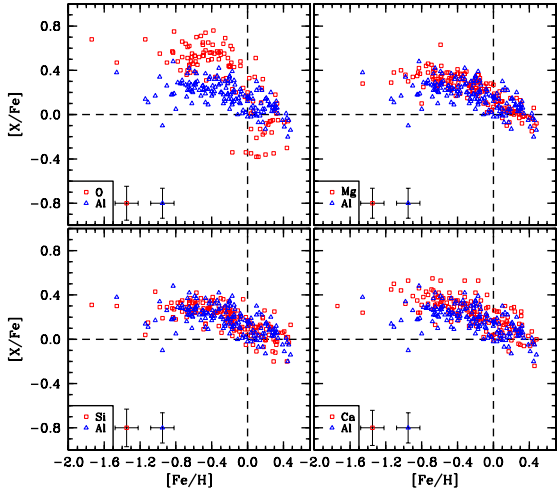}
\caption{[Al/Fe] ratios (open blue triangles) for all bulge and NGC 6553 RGB
stars analyzed here are compared to the abundance trends of the
$\alpha$--elements O, Mg, Si, and Ca (open red boxes).}
\label{f17}
\end{figure}

Given the similar behavior of [Al/Fe] to many of the $\alpha$--elements, in 
Figure \ref{f17} we provide a detailed comparison between [Al/Fe], [O/Fe],
[Mg/Fe], [Si/Fe], and [Ca/Fe] for the bulge stars analyzed here.  While the 
[O/Fe] trend is clearly different than that of [Al/Fe], there are no similarly
strong discrepancies between [Al/Fe] and the other $\alpha$--elements.  At 
[Fe/H]$<$--0.8 both [Mg/Fe] and [Ca/Fe] are $\sim$0.10--0.15 dex enhanced 
compared to [Al/Fe], but those differences disappear at higher [Fe/H].  On
the other hand, the [Si/Fe] and [Al/Fe] trends are essentially identical at 
all [Fe/H] with an average difference of 0.01 dex ($\sigma$=0.13 dex).

Examining the NGC 6553 stars in Figure \ref{f11} shows that Na, and to a lesser
extent Al, exhibit larger star--to--star [Na/Fe] and [Al/Fe] variations than
similar metallicity field stars.  In particular, the average Na and Al 
abundances for the cluster stars are $\langle$[Na/Fe]$\rangle$=$+$0.16 
($\sigma$=0.20) and $\langle$[Al/Fe]$\rangle$=$+$0.17 ($\sigma$=0.13), which
can be compared to the similar metallicity fields stars having 
$\langle$[Na/Fe]$\rangle$=$+$0.03 ($\sigma$=0.11) and 
$\langle$[Al/Fe]$\rangle$=$+$0.16 ($\sigma$=0.10), respectively.  The larger 
[Na/Fe] abundance and dispersion values for the cluster stars suggests NGC 6553
experienced some degree of self--enrichment.  However, unlike low metallicity 
globular clusters, NGC 6553 does not exhibit a strong Na--Al correlation.  This
is in agreement with the observed trend that the Na--Al correlation is more
mild and [Al/Fe] dispersions smaller in metal--rich as opposed to metal--poor 
globular clusters (e.g., Carretta et al. 2009; O'Connell et al. 2011; Cordero 
et al. 2014).  Unfortunately, the 6300 \AA\ telluric oxygen emission feature
combined with NGC 6553's relatively low radial velocity prohibited us from
obtaining an [O/Fe] abundance for more than one star in NGC 6553.  Therefore,
we cannot comment further on the existence or extension of the likely O--Na 
correlation.  Finally, we note that our mean [Na/Fe] and [Al/Fe] values and
abundance dispersions are in excellent agreement with those found by 
Alves--Brito et al. (2006), but are considerably lower than the values (based
on two stars) of Barbuy et al. (1999).

\subsection{The Fe--Peak Elements: Chromium, Cobalt, Nickel, and Copper}

Unlike the lighter elements, the abundance patterns of Fe--peak elements in
the Galactic bulge are not well--explored.  The production of Fe--peak elements
occurs through a variety of processes in the late stages of massive star 
evolution, the resulting core collapse SNe, and also in Type Ia SNe.  The 
Fe--peak abundance patterns can also be useful indicators of a stellar 
population's IMF, with odd--Z elements in particular providing some diagnostic 
power (e.g., Nomoto et al. 2013).  Some initial work on the bulge Fe--peak 
abundance distribution was included in McWilliam \& Rich (1994), which found
[V/Fe], [Cr/Fe], and [Ni/Fe] ratios near solar and a possible enhancement in
[Co/Fe] and [Sc/Fe].  More recent work analyzing the Fe--peak abundance 
trends in the bulge has come from microlensed dwarf studies (Cohen et al. 2008;
Johnson et al. 2008; Cohen et al. 2009; Bensby et al. 2010a; Epstein et al. 
2010; Bensby et al. 2011; Bensby et al. 2013).  The bulge [Mn/Fe] trend in
RGB stars has also been investigated recently by Barbuy et al. (2013).  The
results of the these analyses indicate that the bulge Fe--peak trends are 
similar to that of the local disk, except that the bulge may have different
[Mn/O] ratios than the thick disk for a given [O/H] value.

The general [X/Fe] versus [Fe/H] abundance trends derived here are shown in
Figure \ref{f11}.  From these data we find that: (1) Cr is the element that 
most closely tracks Fe with $\langle$[Cr/Fe]$\rangle$=0.00 ($\sigma$=0.11),
(2) [Co/Fe] exhibits low level variations as a function of [Fe/H] but is 
generally enhanced with $\langle$[Co/Fe]$\rangle$=$+$0.14 ($\sigma$=0.11),
(3) [Ni/Fe] shows similar variations to [Co/Fe] but at a much smaller amplitude
and is slightly enhanced with $\langle$[Ni/Fe]$\rangle$=$+$0.09 
($\sigma$=0.06), (4) the Cu abundance increases monotonically from 
[Cu/Fe]=--0.84 in the most metal--poor star to [Cu/Fe]$\sim$$+$0.40 in the 
most metal--rich stars, and (5) there are no significant Fe--peak abundance 
variations between NGC 6553 stars and the field stars.  

Although the exact nature of Cu nucleosynthesis is complex (e.g., see Mishenina
et al. 2002 and references therein), the significant secondary (i.e., 
metallicity--dependent) production of Cu (and also Na) is evident in 
Figure \ref{f11}.  Additionally, Figure \ref{f12} shows that despite the larger
measurement errors in both O and Cu abundances, the [Cu/O] ratio is strongly
correlated with [Fe/H].  This trend has been noted previously and is prevalent 
in stellar populations with different star formation histories, such as the 
local disk and Sagittarius Dwarf Galaxy (e.g., McWilliam et al. 2013).  The 
[Cu/O] trend is taken as evidence that a significant portion of Cu is 
synthesized in massive stars, perhaps via the weak s--process (e.g., Sneden et 
al. 1991).  However, some component of Cu may also be produced by Type Ia 
SNe (Matteucci et al. 1993).

\begin{figure}
\epsscale{1.0}
\plotone{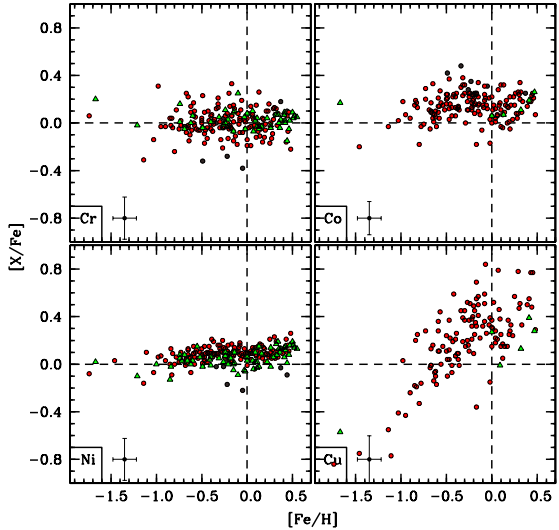}
\caption{Plots comparing the [Cr/Fe], [Co/Fe], [Ni/Fe], and [Cu/Fe] abundances
of the bulge stars measured here (filled red circles) with literature
measurements of bulge microlensed dwarfs (filled green triangles) and field
RGB/red clump stars (filled dark gray circles).  The literature data are from
the sources referenced in Figures \ref{f13} and \ref{f15}.}
\label{f18}
\end{figure}

In Figure \ref{f18} we compare our derived Fe--peak abundance trends with 
those in the literature.  For Cr there is general agreement between
the bulge RGB stars analyzed here and the literature microlensed
dwarf data.  However, the small number of bulge literature data points for 
Co and Cu makes a direct comparison difficult.  The [Ni/Fe] comparison also 
shows excellent agreement overall, but the RGB stars appear systematically 
enhanced by $\la$0.1 dex in the range [Fe/H]=--0.3 to $+$0.1.  Note also the 
similarly small star--to--star dispersion in especially [Ni/Fe] between the RGB
and dwarf data.

\begin{figure}
\epsscale{1.0}
\plotone{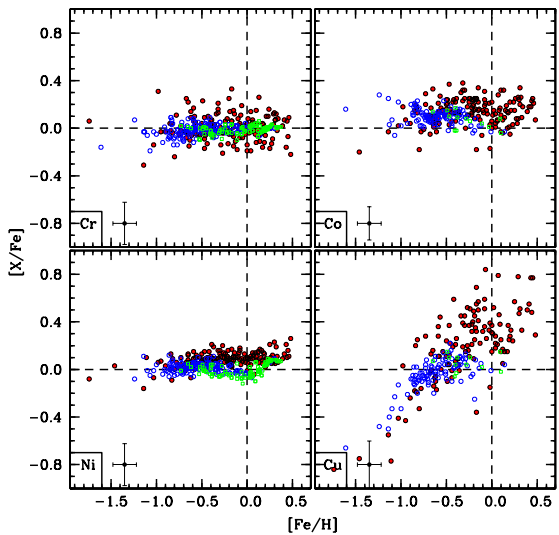}
\caption{Plots comparing the [Cr/Fe], [Co/Fe], [Ni/Fe], and [Cu/Fe] abundances
of the bulge stars measured here (filled red circles) with literature data
for the thick disk (open blue circles) and thin disk (open green boxes).
The literature data are from the sources referenced in Figure \ref{f14}.}
\label{f19}
\end{figure}

A comparison between the bulge Fe--peak abundance trends and those of the
thin/thick disk is shown in Figure \ref{f19}.  Interestingly, at least for 
[Fe/H]$\ga$--1.5, the [Cr/Fe] distribution is seemingly independent of 
formation environment with the bulge, thick disk, and thin disk stars all
having [Cr/Fe]$\sim$0.  For [Co/Fe], [Ni/Fe], and [Cu/Fe], there is significant
overlap between the bulge and thick disk trends at [Fe/H]$\la$--0.5.  At higher
[Fe/H], the bulge may be enhanced in all three elements relative to both the
thick and thin disks.  This is especially evident in the Figure \ref{f19} panel
showing [Ni/Fe] versus [Fe/H]; the low star--to--star scatter in [Ni/Fe] for
all three populations highlights the possible composition difference between
the local disk and bulge from [Fe/H]$\sim$--0.4 to $+$0.2.  While the strong
rise in [Cu/Fe] with metallicity is, as mentioned previously, a common feature 
in many different stellar populations, the bulge stars at [Fe/H]$\ga$--0.3
appear to extend to higher abundances than the local disk.  However, the 
increased measurement uncertainty of Cu and paucity of disk [Cu/Fe] ratios at 
[Fe/H]$>$0 prevents us from undertaking a more comprehensive analysis.

\subsection{Comparing Composition Data to Bulge Chemical Enrichment Models}

\begin{figure}
\epsscale{1.0}
\plotone{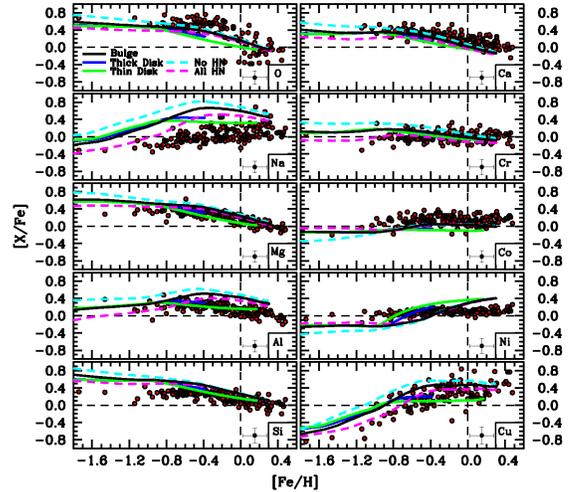}
\caption{Chemical abundance trends are plotted as a function of [Fe/H] and
compared to various chemical enrichment models.  The solid black, blue, and
green lines represent the baseline models from Kobayashi et al. (2006; 2011)
for the Galactic bulge, thick disk, and thin disk, respectively.  The dashed
cyan and magenta lines illustrate how the bulge model changes if the hypernova
fraction is 0 and 1, respectively, for masses $>$20M$_{\rm \odot}$.  Note that
[Ni/Fe] in particular suffers from over--production from Type Ia SNe at
[Fe/H]$>$--1.  Some other elements (e.g., Si) may also be better fit if
systematic offsets were applied.}
\label{f20}
\end{figure}

\begin{figure}
\epsscale{1.0}
\plotone{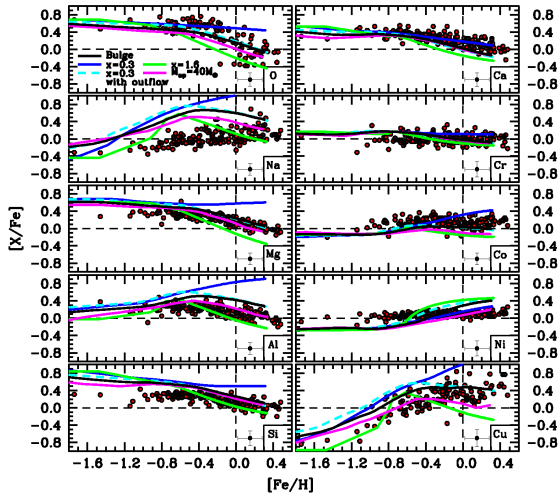}
\caption{Similar to Figure \ref{f20}, the solid black line is our adopted
baseline bulge model from Kobayashi et al. (2006; 2011).  The solid blue line
is the baseline bulge model with a top--heavy (flatter) IMF, and the dashed
cyan line is the same model but with outflow and an increased Type Ia SN rate
(10$\times$).  The solid green line is the baseline bulge model with a steep
IMF.  The solid magenta line is the baseline bulge model with the IMF
truncated at an upper mass limit of 40 M$_{\rm \odot}$.}
\label{f21}
\end{figure}

Accurately modeling the chemical enrichment history of a stellar system 
requires solving for a variety of free parameters that may include the IMF, 
star formation rate, star formation efficiency, supernova/hypernova 
ratio\footnote{Note that the model hypernova fractions only affect stars with
M$>$20 M$_{\rm \odot}$}, inflow/outflow rate, binary fraction, stellar 
evolution time scales, mass loss rates, and stellar yields.  While not all of 
the required input parameters are yet well--defined based on observed data, 
chemical enrichment models are effective tools for examining and interpreting 
chemical composition data.  Therefore, in Figures \ref{f20}--\ref{f21} we 
compare our derived abundance trends with those predicted by chemical 
enrichment models in which parameters such as the IMF, binary fraction, 
supernova/hypernova ratio, and outflow rate are varied.

The baseline Galactic bulge model shown in Figures \ref{f20}--\ref{f21} is 
from Kobayashi et al. (2006; 2011) and is designed to reproduce the metallicity
distribution in Baade's Window from Zoccali et al. (2008), assumes a Kroupa 
(2008) IMF, and assumes a star formation time scale of 3 Gyr (see Kobayashi et 
al. 2011; their Table 1 and Section 2.4 for more detail regarding model input 
parameters).  In general, the baseline model does a reasonable job of 
reproducing the observed abundance trends of all abundance 
ratios, except [Na/Fe] and [Al/Fe].  All of the models shown in 
Figures \ref{f20}--\ref{f21} predict large over--abundances of both
[Na/Fe] and [Al/Fe] that are not observed, suggesting the massive star 
yields of both elements may be too high\footnote{Noting again the possible
effects of additional physics in the stellar models, adding rotation would
likely increase the Na and Al yields.}.  However, as can be seen in 
Figure \ref{f20}, the enhanced Fe production from hypernovae (HNe) decreases 
the [Na/Fe] and [Al/Fe] yields and brings the baseline bulge model into better 
agreement with the light element data.  The addition of HNe also provides 
better agreement between the models and observations for the Fe--peak 
elements, with a trade off of [$\alpha$/Fe] ratios that may be slightly too 
low.  In contrast, Figure \ref{f20} also shows that a paucity of HNe generally 
leads to [X/Fe] ratios that are too high.  It seems that a significant 
fraction of HNe are required to accurately reproduce the observed bulge 
abundance trends.  Unfortunately, the HN fraction is best constrained at
[Fe/H]$\la$--1, where data are scarce.

In Figure \ref{f21} we examine how changes in the IMF could affect the expected 
abundance trends.  Compared to the Kroupa (2008) IMF adopted in our baseline
bulge model, a steep IMF (x=1.6) is completely ruled out by the data.  
Additionally, adopting a Kroupa (2008) IMF that truncates at 
40 M$_{\rm \odot}$, and thus ignores contributions from the most massive stars,
is inconsistent with the [Cu/Fe] abundances, and to a lesser extent those of
[Co/Fe].  While a flatter, top--heavy IMF (x=0.3) alone leads to [X/Fe] ratios 
that are too high for nearly every element, a reduction in the yields from 
outflow and/or slow star--formation combined with a high Type Ia SN rate,
artificially enhanced with a ten times larger binary fraction, could bring such 
a model into agreement with the data.  However, bulge formation 
models with slow star--formation are likely unrealistic, and the observed 
[Co/Fe] and [Cu/Fe] data appear to rule out these models.  Based on the present
data it does not appear that the bulge required a uniquely ``non--standard" IMF
to reach its present--day composition (but see also Ballero et al. 2007, for 
example).

Finally, in Figure \ref{f20} we also compare the measured bulge abundance 
trends with our adopted baseline model and similar models representing the 
composition distributions of the local thick disk and thin disk.  Comparing
the three predicted trends indicates that in the range 
--0.8$\la$[Fe/H]$\la$--0.3 the bulge [$\alpha$/Fe] ratios should be similar 
or modestly enhanced and remain enhanced to higher [Fe/H] than the thick disk.  Similarly, at [Fe/H]$\ga$0 the bulge and thin disk should exhibit similar, if 
not identical, [$\alpha$/Fe] ratios.  Both of these predictions match our 
observations (see Section 4.1).  The predicted enhancements in the bulge for 
[Na/Fe] and [Al/Fe] compared to the local disk are not supported by 
observations, but this could be related to the previously mentioned possible
over--production issues of the adopted stellar yields.  However, in addition
to Na and Al, Figure \ref{f20} shows that Co and Cu may also exhibit some
discriminating power between the bulge and local disk populations.  In 
particular, the data support bulge stars with [Fe/H]$\ga$--0.5 having [Co/Fe] 
and [Cu/Fe] ratios that are higher than the local disk.  Therefore, the data
and models presented here provide some supporting evidence that the bulge
experienced a different chemical enrichment path than the thick disk.

\section{SUMMARY}

We have measured radial velocities and chemical abundances of O, Na, Mg, Al,
Si, Ca, Cr, Fe, Co, Ni, and Cu in a sample of 156 RGB stars located in
Galactic bulge fields centered near (l,b)=($+$5.25,--3.02) and (0,--12).  The
($+$5.25,--3.02) also includes 12 stars identified as likely members of 
the bulge globular cluster NGC 6553, based on their radial velocity and [Fe/H]
values.  The results are based on high resolution archival spectra obtained 
with the FLAMES--GIRAFFE instrument, and originally used to derive [Fe/H] and 
[$\alpha$/Fe] abundances in Zoccali et al. (2008) and Gonzalez et al. (2011).  
We culled the original target list and selected only those stars with co--added
S/N$\ga$70 that also lack strong TiO bands.  The abundance analysis was 
carried--out using standard EW and spectrum synthesis techniques.

Our derived heliocentric radial velocity distributions for both fields are 
in good agreement with past surveys (BRAVA, GIBS, and APOGEE) covering nearby 
fields.  We do not confirm the existence of a significant population of high 
velocity stars noted by Nidever et al. (2012) and Babusiaux et al. (2014).  
However, our targeted fields are farther away from the plane than most of 
those in which Nidever et al. (2012) and Babusiaux et al. (2014) observe the 
cold, high velocity stars.  For both fields analyzed here we also 
find that the velocity dispersion monotonically decreases with increasing 
[Fe/H].  This is not unexpected for the outer bulge field at (0,--12), but the
similar trend in the ($+$5.25,--3.02) field appears to contradict the findings 
of Babusiaux et al. (2010; 2014) that the velocity dispersion of bulge stars 
with [Fe/H]$\ga$0 increases at lower Galactic latitude.  The reason for this 
discrepancy is not clear, but we note that previous analyses finding increased 
velocity dispersion at low Galactic latitude for metal--rich stars have all 
focused on minor--axis fields.  The inner bulge field included here is several 
degrees off--axis.

The composition data reconfirm the already well--documented metallicity 
gradient in the bulge.  Similarly, we find good agreement between our derived
[Mg/Fe], [Si/Fe], and [Ca/Fe] abundances and those of Gonzalez et al. (2011).
Additionally, we confirm that there are no significant field--to--field
[$\alpha$/Fe] abundance variations among various bulge sight lines.  Our new
$\alpha$--element measurements also reinforce the previously held notion
(e.g., McWilliam et al. 2008) that the decline in [O/Mg] with increasing 
metallicity is likely the result of metallicity dependent yields from massive
stars.  While we find that the bulge and thick disk exhibit nearly identical
[$\alpha$/Fe] ratios at [Fe/H]$\la$--0.5, the bulge stars appear to remain
enhanced in [$\alpha$/Fe] by up to 0.1--0.2 dex higher in [Fe/H] than the 
local thick disk.  The bulge [$\alpha$/Fe] ratios at [Fe/H]$\ga$0 are 
well--matched to the local thin disk trends.  These results are in agreement 
with recent differential abundance analyses of microlensed bulge dwarfs (Bensby
et al. 2013), and suggest the bulge experienced faster enrichment than the 
local thick disk.  However, similar differential analyses comparing bulge and 
thick disk giants find no significant differences between the two populations 
(Mel{\'e}ndez et al. 2008; Alves--Brito et al. 2010; Gonzalez et al. 2011).

Combining the new data set of [$\alpha$/Fe] abundances with those available
in the literature now totals several hundred stars.  However, the combined
data set does not reveal any significant population with ``anomalous"
chemistry, such as the low [$\alpha$/Fe] ratios reminiscent of many 
present--day dwarf galaxy stars.  Therefore, we can effectively rule out these
types of objects as major contributors to any portion of the present--day 
Galactic bulge field population.  This further supports the idea that the 
Galactic bulge is not a merger--built system.  Similarly, the [$\alpha$/Fe] 
ratios of the NGC 6553 stars are identical to those of similar metallicity
field stars.  This suggests NGC 6553 formed \emph{in situ} with the bulge and
is not a captured system.

With regard to the light, odd--Z elements, we find that Na and Al exhibit 
discrepant trends as a function of metallicity.  In particular, bulge stars
exhibit a steady increase in [Na/Fe] with increasing [Fe/H], but the [Al/Fe]
trend almost exactly matches that of the $\alpha$--elements (except oxygen).
While we do not find any significant field--to--field variations in either 
[Na/Fe] or [Al/Fe], our results indicate that the bulge and thick disk have 
different [Na/Fe] abundances at [Fe/H]$\la$--0.5 but similar [Al/Fe].  
Interestingly, the ``$\alpha$--like" behavior of [Al/Fe] contrasts with several
previous bulge studies that found [Al/Fe] was enhanced up to [Fe/H]=$+$0.5.
Instead, our results are in agreement with the abundance patterns of
microlensed bulge dwarfs (e.g., Bensby et al. 2013).  The discrepant behavior
of Na and Al suggests metallicity dependent yields from massive stars, and 
perhaps intermediate mass stars, leads to significantly more production of 
Na than Al at high metallicity.  We also find that the NGC 6553 stars have 
nearly identical [Al/Fe] ratios as similar metallicity field stars, but both 
the average [Na/Fe] abundance and star--to--star dispersion of cluster stars 
are higher.  This suggests NGC 6553 experienced some light element 
self--enrichment, which is typical for globular clusters.

The abundance trends of the Fe--peak elements are distinctly different: (1) the
average [Cr/Fe] ratio is essentially solar over the full range in [Fe/H]
and shows no variations over the metallicity range probed here, (2) both
[Co/Fe] and [Ni/Fe] are enhanced by $\sim$$+$0.1 dex at nearly all [Fe/H] and 
exhibit some low amplitude, metallicity--dependent variations, and (3) [Cu/Fe]
exhibits a large increase from the metal--poor to metal--rich end of the 
distribution.  In a similar fashion to [Na/Fe], the strong secondary 
(metallicity--dependent) production of Cu is evident in bulge stars, and the 
correlation between [Cu/O] and [Fe/H] suggests massive stars produce 
significant portions of Cu.  However, Cu production from another source 
(e.g., Type Ia SNe) seems required to explain the high [Cu/Fe] abundances at
super--solar metallicities.  Interestingly, at [Fe/H]$\ga$--2 the [Cr/Fe] 
trend is identical between the bulge, thick disk, and thin disk, but the 
heavier Fe--peak [X/Fe] ratios appear to all be enhanced in the bulge relative 
to the local disk.  Additionally, the NGC 6553 Fe--peak abundance trends are 
in agreement with similar metallicity field stars.

Despite predicting [Na/Fe], [Al/Fe], and [Ni/Fe] ratios that are too high, our 
adopted baseline bulge chemical enrichment model from Kobayashi et al. (2006;
2011) does a reasonable job fitting the abundance trends of the $\alpha$ and 
other Fe--peak elements.  However, better agreement between the data and model
is found when a significant fraction of HNe, which produce more Fe, are 
included.  Unfortunately, setting the HN fraction is best 
constrained using abundance patterns at [Fe/H]$\la$--1, where the bulge data
are sparse.  While a Kroupa (2008) IMF provides a reasonable fit to the 
observed abundance trends, a top--heavy IMF including strong outflow cannot be 
ruled out.  In contrast, the Fe--peak abundance data strongly rule out IMFs
that are truncated to exclude the contributions of stars $>$40 M$_{\rm \odot}$,
steep IMFs (e.g., x=1.6), and top--heavy IMFs that do not include outflow.
We conclude that the bulge likely does not require a particularly unusual IMF
to explain its present--day abundance patterns, and that its enhanced 
abundances for several $\alpha$ and Fe--peak elements match model predictions
in which the bulge experienced a different enrichment history than the local
disk.

\acknowledgements

We thank the anonymous referee for a careful reading of the manuscript and 
helpful comments that lead to improvement of the manuscript.  This research 
has made use of NASA's Astrophysics Data System Bibliographic Services.  This 
publication makes use of data products from the Two Micron All Sky Survey, 
which is a joint project of the University of Massachusetts and the Infrared 
Processing and Analysis Center/California Institute of Technology, funded by 
the National Aeronautics and Space Administration and the National Science 
Foundation.  CIJ gratefully acknowledges support from the Clay Fellowship, 
administered by the Smithsonian Astrophysical Observatory.  RMR acknowledges 
support from NSF-AST-1212095.  AK acknowledges the Deutsche 
Forschungsgemeinschaft for funding from Emmy--Noether grant Ko 4161/1.

\clearpage
\LongTables
\begin{landscape}

\tablenum{1}
\tablecolumns{15}
\tablewidth{0pt}

\setlength{\hoffset}{-0.90in}
\begin{deluxetable}{ccccccccccccccc}
\tabletypesize{\scriptsize}
\tablecaption{Star Identifications, Coordinates, Model Atmosphere
Parameters, and Radial Velocities}
\tablehead{
\colhead{ID\tablenotemark{a}}   &
\colhead{2MASS} &
\colhead{RA (J2000)}      &
\colhead{DEC (J2000)}      &
\colhead{V}      &
\colhead{V--I}      &
\colhead{J}      &
\colhead{H}      &
\colhead{K$_{\rm S}$}      &
\colhead{T$_{\rm eff}$}      &
\colhead{log(g)}      &
\colhead{[Fe/H]}      &
\colhead{vt}      &
\colhead{RV$_{\rm helio.}$}      &
\colhead{RV$_{\rm error}$\tablenotemark{b}}      \\
\colhead{}      &
\colhead{}      &
\colhead{(degrees)}      &
\colhead{(degrees)}      &
\colhead{(mag.)}      &
\colhead{(mag.)}      &
\colhead{(mag.)}      &
\colhead{(mag.)}      &
\colhead{(mag.)}      &
\colhead{(K)}      &
\colhead{(cgs)}      &
\colhead{}      &
\colhead{(km s$^{\rm -1}$)}      &
\colhead{(km s$^{\rm -1}$)}      &
\colhead{(km s$^{\rm -1}$)}
}

\startdata
\hline
\multicolumn{15}{c}{(l,b)=($+$5.25,--3.02)}       \\
\hline
119799C4        &       18081026$-$2547354      &       272.042708      &       $-$25.793167    &       \nodata &       \nodata &       12.856  &       11.986  &       11.704  &       4400    &       2.05    &       $-$0.39 &       1.90    &       $+$57.96        &       0.11    \\
129499C4        &       18081617$-$2543192      &       272.067375      &       $-$25.721972    &       \nodata &       \nodata &       12.596  &       11.836  &       11.608  &       4900    &       2.75    &       $+$0.13 &       2.10    &       $+$27.10        &       0.04    \\
176772C5        &       18080936$-$2556168      &       272.039417      &       $-$25.938028    &       \nodata &       \nodata &       11.710  &       10.794  &       10.483  &       4225    &       2.15    &       $+$0.06 &       1.90    &       $+$37.82        &       0.08    \\
181349C5        &       18080465$-$2554356      &       272.019292      &       $-$25.909917    &       \nodata &       \nodata &       12.924  &       12.253  &       11.906  &       4775    &       2.75    &       $+$0.26 &       2.10    &       $+$38.51        &       0.04    \\
183783C5        &       18080912$-$2553420      &       272.037958      &       $-$25.895028    &       \nodata &       \nodata &       12.876  &       12.119  &       11.848  &       4700    &       2.75    &       $+$0.10 &       1.75    &       $-$97.45        &       0.01    \\
184088C5        &       18080726$-$2553356      &       272.030250      &       $-$25.893222    &       \nodata &       \nodata &       12.789  &       11.869  &       11.640  &       4500    &       2.50    &       $-$0.47 &       1.80    &       $+$173.45       &       0.22    \\
\enddata

\tablenotetext{a}{These are the OGLE identification values given in the
image headers, and are also listed in Zoccali et al. (2008) and Gonzalez et
al. (2011).}
\tablenotetext{b}{The radial velocity errors represent the 1$\sigma$ values
from individual exposures of each star over all filters.}
\end{deluxetable}
\clearpage
\end{landscape}

\clearpage
\tablenum{2}
\tablecolumns{7}
\tablewidth{0pt}

\begin{deluxetable}{ccccccc}
\tablecaption{Line List}
\tablehead{
\colhead{Species}       &
\colhead{Wavelength}    &
\colhead{E.P.}  &
\colhead{log(gf)\tablenotemark{a}}        &
\colhead{log $\epsilon$(X)$_{\rm \odot}$}       &
\colhead{log $\epsilon$(X)$_{Arc.}$}    &
\colhead{[X/Fe] or [Fe/H]$_{Arc.}$}     \\
\colhead{}      &
\colhead{(\AA)} &
\colhead{(eV)}  &
\colhead{}      &
\colhead{}      &
\colhead{}      &
\colhead{}
}

\startdata
[O I]   &       6300.30 &       0.00    &       $-$9.750        &       8.69    &       8.63    &       $+$0.44 \\
Na I    &       6154.23 &       2.10    &       $-$1.560        &       6.33    &       5.89    &       $+$0.06 \\
Na I    &       6160.75 &       2.10    &       $-$1.210        &       6.33    &       5.89    &       $+$0.06 \\
Mg I    &       6318.71 &       5.10    &       $-$2.010        &       7.58    &       7.38    &       $+$0.30 \\
Mg I    &       6319.24 &       5.10    &       $-$2.250        &       7.58    &       7.38    &       $+$0.30 \\
Mg I    &       6319.49 &       5.10    &       $-$2.730        &       7.58    &       7.38    &       $+$0.30 \\
Al I    &       6696.02 &       3.14    &       $-$1.570        &       6.47    &       6.28    &       $+$0.31 \\
Al I    &       6698.67 &       3.14    &       $-$1.890        &       6.47    &       6.28    &       $+$0.31 \\
\enddata

\tablenotetext{a}{The ``hfs" designation indicates the abundance was
calculated taking hyperfine structure into account.  See text for details.}
\end{deluxetable}

\clearpage
\begin{landscape}
\tablenum{3}
\tablecolumns{13}
\tablewidth{0pt}

\begin{deluxetable}{ccccccccccccc}
\tabletypesize{\scriptsize}
\tablecaption{Abundance Ratios}
\tablehead{
\colhead{ID}    &
\colhead{[O/Fe]\tablenotemark{a}}       &
\colhead{[Na/Fe]}      &
\colhead{[Mg/Fe]}      &
\colhead{[Al/Fe]}      &
\colhead{[Si/Fe]}      &
\colhead{[Ca/Fe]}      &
\colhead{[Cr/Fe]}      &
\colhead{[Fe I/H]}      &
\colhead{[Fe II/H]}      &
\colhead{[Co/Fe]}      &
\colhead{[Ni/Fe]}      &
\colhead{[Cu/Fe]}
}

\startdata
\hline
\multicolumn{13}{c}{(l,b)=($+$5.25,--3.02)}       \\
\hline
119799C4        &       $+$0.63 &       $+$0.08 &       $+$0.29 &       $+$0.26 &       $+$0.38 &       $+$0.17 &       $+$0.07 &       $-$0.39 &       $-$0.38 &       $+$0.25 &       $+$0.21 &       $+$0.26 \\
129499C4        &       \nodata &       $+$0.25 &       $+$0.22 &       $+$0.24 &       $+$0.04 &       $+$0.30 &       $+$0.26 &       $+$0.13 &       \nodata &       $+$0.32 &       $+$0.06 &       \nodata \\
176772C5        &       $+$0.12 &       $+$0.19 &       $+$0.15 &       $+$0.08 &       $-$0.05 &       $+$0.16 &       $+$0.05 &       $+$0.07 &       $+$0.05 &       $+$0.07 &       $+$0.14 &       \nodata \\
181349C5        &       $-$0.06 &       $+$0.58 &       $+$0.15 &       $+$0.20 &       $+$0.11 &       $+$0.08 &       $+$0.00 &       $+$0.30 &       $+$0.21 &       $+$0.25 &       $+$0.10 &       \nodata \\
183783C5        &       \nodata &       $+$0.32 &       $+$0.12 &       $+$0.30 &       $-$0.11 &       $+$0.15 &       $+$0.15 &       $+$0.10 &       \nodata &       $+$0.32 &       $+$0.08 &       \nodata \\
184088C5        &       $+$0.74 &       $+$0.03 &       $+$0.30 &       $+$0.37 &       $+$0.14 &       $+$0.37 &       $+$0.09 &       $-$0.45 &       $-$0.48 &       $+$0.37 &       $+$0.15 &       $+$0.37 \\
184618C5        &       \nodata &       $+$0.00 &       $+$0.63 &       $+$0.28 &       $+$0.24 &       $+$0.31 &       $+$0.06 &       $-$0.58 &       $-$0.59 &       $+$0.14 &       $+$0.15 &       $+$0.28 \\
\enddata

\tablenotetext{a}{The [O/Fe] ratio is normalized to the [Fe II/H] abundance.
For stars without a [Fe II/H] measurement, the [O/Fe] ratio was normalized to
the [Fe I/H] abundance.}
\end{deluxetable}

\clearpage
\end{landscape}

\clearpage
\begin{landscape}
\tablenum{4}
\tablecolumns{13}
\tablewidth{0pt}

\begin{deluxetable}{ccccccccccccc}
\tabletypesize{\scriptsize}
\tablecaption{Total Abundance Uncertainty for $\Delta$T$_{\rm eff}$$+$100 K;
$\Delta$log(g)$+$0.3 cgs; $\Delta$[M/H]$+$0.15 dex; $\Delta$vt$+$0.3 km
s$^{\rm -1}$}
\tablehead{
\colhead{ID}    &
\colhead{$\Delta$[O/Fe]}   &
\colhead{$\Delta$[Na/Fe]}   &
\colhead{$\Delta$[Mg/Fe]}   &
\colhead{$\Delta$[Al/Fe]}   &
\colhead{$\Delta$[Si/Fe]}   &
\colhead{$\Delta$[Ca/Fe]}   &
\colhead{$\Delta$[Cr/Fe]}   &
\colhead{$\Delta$[Fe I/H]}      &
\colhead{$\Delta$[Fe II/H]}     &
\colhead{$\Delta$[Co/Fe]}   &
\colhead{$\Delta$[Ni/Fe]}   &
\colhead{$\Delta$[Cu/Fe]}
}

\startdata
\hline
\multicolumn{13}{c}{(l,b)=($+$5.25,--3.02)}       \\
\hline
119799C4        &       0.19    &       0.08    &       0.03    &       0.08    &       0.10    &       0.14    &       0.09    &       0.11    &       0.27    &       0.05    &       0.05    &       0.15    \\
129499C4        &       \nodata &       0.06    &       0.04    &       0.05    &       0.09    &       0.09    &       0.05    &       0.10    &       \nodata &       0.05    &       0.04    &       \nodata \\
176772C5        &       0.20    &       0.12    &       0.04    &       0.10    &       0.08    &       0.17    &       0.09    &       0.13    &       0.25    &       0.04    &       0.03    &       \nodata \\
181349C5        &       0.14    &       0.10    &       0.03    &       0.07    &       0.08    &       0.12    &       0.07    &       0.11    &       0.21    &       0.05    &       0.04    &       \nodata \\
183783C5        &       \nodata &       0.09    &       0.04    &       0.07    &       0.08    &       0.12    &       0.07    &       0.10    &       \nodata &       0.05    &       0.04    &       \nodata \\
184088C5        &       0.17    &       0.09    &       0.03    &       0.07    &       0.09    &       0.14    &       0.09    &       0.10    &       0.27    &       0.05    &       0.05    &       0.15    \\
184618C5        &       \nodata &       0.06    &       0.03    &       0.05    &       0.08    &       0.10    &       0.07    &       0.10    &       0.20    &       0.06    &       0.04    &       0.15    \\
\enddata

\end{deluxetable}

\clearpage
\end{landscape}

\clearpage
\begin{landscape}
\tablenum{5}
\tablecolumns{13}
\tablewidth{0pt}

\begin{deluxetable}{ccccccccccccc}
\tabletypesize{\scriptsize}
\tablecaption{1$\sigma$ Line--to--Line Abundance Dispersion}
\tablehead{
\colhead{ID}    &
\colhead{$\sigma$$_{\rm O}$}    &
\colhead{$\sigma$$_{\rm Na}$}   &
\colhead{$\sigma$$_{\rm Mg}$}   &
\colhead{$\sigma$$_{\rm Al}$}   &
\colhead{$\sigma$$_{\rm Si}$}   &
\colhead{$\sigma$$_{\rm Ca}$}   &
\colhead{$\sigma$$_{\rm Cr}$}   &
\colhead{$\sigma$$_{\rm Fe I}$}   &
\colhead{$\sigma$$_{\rm Fe II}$}   &
\colhead{$\sigma$$_{\rm Co}$}   &
\colhead{$\sigma$$_{\rm Ni}$}   &
\colhead{$\sigma$$_{\rm Cu}$}
}

\startdata
\hline
\multicolumn{13}{c}{(l,b)=($+$5.25,--3.02)}       \\
\hline
119799C4        &       0.08    &       0.02    &       0.00    &       0.02    &       0.12    &       0.04    &       0.17    &       0.13    &       0.08    &       0.09    &       0.10    &       0.08    \\
129499C4        &        \nodata        &       0.04    &       0.07    &       0.11    &       0.17    &       0.12    &       0.10    &       0.14    &           \nodata     &       0.04    &       0.18    &       \nodata \\
176772C5        &       0.08    &       0.01    &       0.04    &       0.04    &       0.15    &       0.15    &       0.08    &       0.17    &       0.08    &       0.05    &       0.16    &       \nodata \\
181349C5        &       0.08    &       0.25    &       0.07    &       0.08    &       0.21    &       0.08    &       0.18    &       0.15    &       0.08    &       0.07    &       0.16    &       \nodata \\
183783C5        &        \nodata        &        \nodata        &       0.10    &       0.08    &       0.06    &       0.11    &       0.22    &       0.13    &           \nodata     &       0.08    &       0.18    &          \nodata      \\
184088C5        &       0.08    &       0.00    &       0.07    &       0.00    &       0.11    &       0.09    &       0.16    &       0.13    &       0.00    &       0.04    &       0.12    &       0.08    \\
184618C5        &        \nodata        &       0.11    &       0.07    &       0.00    &       0.15    &       0.10    &       0.13    &       0.14    &       0.08    &       0.08    &       0.10    &       0.08    \\
\enddata
\tablecomments{For cases where only one line was measured we have assigned a
default value of 0.08.  This is the average dispersion value for all species
in which more than one line was measured.}

\end{deluxetable}
\clearpage
\end{landscape}

\end{document}